\documentclass[aps,twocolumn,pra,superscriptaddress,showpacs,showkeys,floatfix]{revtex4-1}
\usepackage{graphicx,color}% Include figure files
\usepackage{epsfig}
\usepackage{graphicx}
\usepackage{amsmath,amssymb}
\usepackage{mathrsfs}
\usepackage{color}
\usepackage{bm}%

\def\Tr{\textrm{Tr}~}

 \newcommand{\bra}[1]{\langle{#1} |}
 \newcommand{\ket}[1]{|{#1}\rangle  }
 
 \newcommand{\ketbra}[2]{\vert {#1} \rangle \langle{#2}\vert}

\providecommand{\openone}{\leavevmode\hbox{\small1\kern-3.8pt\normalsize1}}

\begin{document}

\title{Information transmission over an amplitude damping channel with an 
arbitrary degree of memory}

\author{Antonio D'Arrigo}
\affiliation{Dipartimento di Fisica e Astronomia,
Universit\`a degli Studi Catania \& CNR-IMM UOS Universit\`a (MATIS), Consiglio Nazionale delle Ricerche,
Via Santa Sofia 64, 95123 Catania, Italy}
\author{Giuliano Benenti}
\affiliation{Center for Nonlinear and Complex Systems,
Universit\`a degli Studi dell'Insubria, Via Valleggio 11, 22100 Como, Italy}
\affiliation{Istituto Nazionale di Fisica Nucleare, Sezione di Milano,
via Celoria 16, 20133 Milano, Italy}
\author{Giuseppe Falci}
\affiliation{Dipartimento di Fisica e Astronomia,
Universit\`a degli Studi Catania \& CNR-IMM UOS Universit\`a (MATIS), Consiglio Nazionale delle Ricerche,
Via Santa Sofia 64, 95123 Catania, Italy}
\affiliation{Istituto Nazionale di Fisica Nucleare, Sezione di Catania, Via S. Sofia 64, 95123 Catania, Italy}
\author{Chiara Macchiavello}
\affiliation{Dipartimento di Fisica and INFN-Sezione di Pavia, Via Bassi
6, I-27100 Pavia, Italy}

\begin{abstract}  
We study the performance of a partially correlated amplitude damping 
channel acting on two qubits. We derive lower bounds for the single-shot 
classical capacity by studying two kinds of quantum ensembles, one which 
allows to maximize the Holevo quantity for the memoryless channel and the 
other allowing the same task but for the full-memory channel. 
In these two cases, we also show the amount of entanglement 
which is involved in achieving the maximum of the Holevo quantity.
For the single-shot quantum capacity we discuss both a lower and an upper 
bound, achieving a good estimate for high values of the channel 
transmissivity. We finally compute the entanglement-assisted classical channel 
capacity. 
\end{abstract}                                                                 

%Da sistemare

%03.67.-a Quantum information,
%03.67.Hk Quantum communication, 
%03.65.Yz Decoherence; open systems; quantum statistical methods
%03.65.Ud Entanglement and quantum nonlocality

\maketitle

\widetext

\section{Introduction}
\label{sec:intro}

One of the key issues of quantum information is the use of quantum systems to
convey information. Although quantum systems are unavoidably 
affected by noise, reliable transmission is still possible by proper coding 
\cite{cover-thomas,nielsen-chuang,benenti-casati-strini,wilde}. Coding involves multiple
channel uses. 
The relevant quantities for classical and quantum information 
transmission are the 
\emph{classical capacity}~\cite{hausladen,schumacher-westmoreland,holevo98} $C$ 
and the \emph{quantum capacity}~\cite{lloyd,barnum,devetak} $Q$, 
defined as the maximum number of,
respectively, bits and and qubits that can be reliably transmitted per
channel use. Finally, the 
\emph{entanglement-assisted classical capacity}~\cite{adami-Cerf,bennett1999,bennett-shor}
$C_E$ is the capacity of transmitting classical information, provided the sender
and the receiver share unlimited prior entanglement. This latter 
quantity is important since it upper bounds the previous ones. We have
$Q\le C\le C_E$. 
The computation of capacities $C$ and $Q$ is in general a hard task,
since a ``regularization'' procedure is requested, namely 
an optimization over all possible 
$n$-use input states, in the limit $n\to\infty$.

In the simplest setting
each channel use is independent of the previous ones. It means that, 
if a \emph{quantum channel} use is described by the map $\mathcal E$,
$n$ uses of the channel are described by the map $\mathcal{E}_n= \mathcal{E}^{\otimes n}$.
This assumption is not always justified. For instance, with increasing 
the transmission rate, the environment may retain \emph{memory} of the previous
channel uses. In this case noise 
introduces memory (or \emph{correlation}) effects among consecutive
channel uses, and $\mathcal{E}_n \neq \mathcal{E}^{\otimes n}$
(\emph{memory channels}). 
Such effects can be investigated experimentally in optical 
fibers~\cite{banaszek} or in solid-state implementations of quantum hardware, affected
by low-frequency noise~\cite{solid-state}.  
Quantum memory channels attracted growing interest in the last years, 
and interesting new features emerged thanks to modeling of 
relevant physical examples, including
depolarizing channels~\cite{mp02,MMM},
Pauli channels~\cite{mpv04,daems,dc}, dephasing
channels~\cite{hamada,njp,virmani,gabriela,lidar},
Gaussian channels~\cite{cerf},
lossy bosonic channels~\cite{mancini,lupo},
spin chains~\cite{spins}, collision models~\cite{collision},
complex network dynamics~\cite{caruso},
and a micromaser model~\cite{micromaser}.
For a recent review on quantum channels with memory effects,
see Ref.~\cite{memo_review}.

Here we study the behavior of a two-qubit 
memory amplitude damping channel. 
We extend
the model introduced in Ref.~\cite{MADC2013} by addressing the cases of partial memory. 
We use a memory parameter $\mu$ which spans from
zero to one allowing us to recover the memoryless case ($\mu=0$) as well as the
full memory case ($\mu=1$). 
We study the channel capability to transmit both 
classical
%~\cite{hausladen,schumacher-westmoreland,holevo98} 
and quantum information
%~\cite{lloyd,barnum,devetak},
as well as the entanglement-assisted classical 
information.
%~\cite{adami-Cerf,bennett1999,bennett-shor}. 
We derive lower bounds for the classical
capacity, lower and upper bounds for the quantum capacity, and
compute the channel capacity
for entanglement-assisted classical communication. 
In all cases we analytically indentify a general 
form of the ensembles that optimize the channel capacities.
Then we perform numerical optimizations for single use of 
the channel, thus deriving lower bounds for $Q$ and $C$, as
well as computing $C_E$, for which the regularization 
$n\to\infty$ is not needed. 
For such ensembles, we also show the populations of the density operators
which solve the optimization problems. Such information 
may provide useful indications for real 
(few channel uses) coding strategies. In the case of the classical
capacity, we investigate two classes of ensembles; we find that neither of them 
is useful to overcome -for the memoryless setting- the limit of
the product state classical capacity of the (memoryless) amplitude 
damping channel~\cite{giovannetti,hastings}.
Finally, we find that any finite amount of memory increases
the amount of reliably transmitted information with respect to the memoryless
case, for all the scenarios considered. 

The paper is organized as follows. In Sec. \ref{sec:model} we describe 
the channel model and the channel covariance
properties. In Sec. \ref{sec:classical-capacity} we study 
the classical capacity of the quantum channel, addressing
the ensembles classes which maximize the Holevo quantity, showing two
distinct lower bounds for the classical capacity.
In Sec. \ref{sec:quantum-capacity} we compute both a lower and an upper
bound for the quantum capacity, which are very close to each other
for good quality (relatively high transmissivity) channels.
In \ref{sec:classical-Entanglement-Assisted Capacity}  we determine the 
quantum capacity and
the entanglement-assisted channel capacity. 
We finish with concluding remarks in Sec.~\ref{sec:conclusions}.

\section{The Model and its covariance properties}
\label{sec:model}

We will first briefly review the memoryless amplitude damping 
channel (\textit{ad})~\cite{nielsen-chuang,benenti-casati-strini}, which 
acts on a generic single-qubit state $\rho$ as follows:
\begin{equation}
 \rho \quad \rightarrow \quad \rho'={\cal E}(\rho)\,
=\,\sum_{i \in \{0,1\}} E_i \,\rho\,E_i^\dag,
 \label{eq:ampl-damping-channel}
\end{equation}
where the Kraus operators $E_i$ are given by
\begin{equation}
{E}_0 = \left(
\begin{array}{cc}
  1 & 0   \\
  0 & \sqrt{\eta}  \\
  \end{array} \right), \qquad
{E}_1 = \left(
\begin{array}{cccc}
    0 & \sqrt{1-\eta}  \\
    0 & 0  \\
  \end{array} \right).
\label{eq:Memoryless-Kraus-Operators}
\end{equation}
Here we are using the orthonormal basis $\{\ket{0},\ket{1}\}$ 
($\sigma_z=\ketbra{0}{0}-\ketbra{1}{1}$). 
This channel describes relaxation processes, such as spontaneous 
emission of an atom, in which the system decays from the excited 
state $\ket{1}$ to the ground state $\ket{0}$. 
The channel acts as follows on a generic single-qubit state:
\begin{equation}
  \rho= 
    \left(
      \begin{array}{cc}
        1-p & \gamma   \\
        \gamma* & p  \\
      \end{array} 
    \right) \quad \rightarrow \quad
  \rho'= {\cal E}(\rho)=
    \left(
      \begin{array}{cc}
        1-\eta \,p & \sqrt{\eta}\,\gamma   \\
        \sqrt{\eta}\, \gamma* & \eta\, p  \\
      \end{array} 
    \right). 
\end{equation}
Note that the noise parameter $\eta$ ($0\le \eta\le 1$) plays the role of 
channel transmissivity. 
Indeed for $\eta=1$ we have a noiseless channel,
whereas for $\eta=0$ the channel cannot carry any information since for any 
possible input we always obtain the same output state $\ket{0}$.

For two memoryless uses we have that
\begin{equation}
 \rho \quad \rightarrow \quad \rho'={\cal E}_0(\rho)\,=\,\sum_{i \in \{0,3\}} A_i \,\rho\,A_i^\dag,
 \label{eq:memoryless-channel}
\end{equation}
where $\rho$ is the density matrix related to a two-qubit system, and 
${\cal E}_0={\cal E}\otimes{\cal E}$ so that
the Kraus operators $A_i$ are given by
\footnotesize
\begin{eqnarray}
 \hspace{-0.2cm}A_0 = E_0 \otimes E_0 = \left(
  \begin{array}{cccc}
  1 & 0 & 0 & 0  \\
  0 & \sqrt{\eta} & 0 & 0  \\
  0 & 0 & \sqrt{\eta} & 0  \\
  0 & 0 & 0 & \eta  \\
  \end{array} \right), \nonumber
\end{eqnarray}
\begin{eqnarray}
 &&A_1 = E_0 \otimes E_1 = \left(
  \begin{array}{cccc}
  0 & \sqrt{1-\eta} & 0 & 0  \\
  0 & 0 & 0 & 0  \\
  0 & 0 & 0 & \sqrt{\eta(1-\eta)}  \\
  0 & 0 & 0 & 0  \\
  \end{array} \right), \nonumber
\end{eqnarray}
\begin{eqnarray}
 &&A_2 = E_1 \otimes E_0 = \left(
  \begin{array}{cccc}
  0 & 0 & \sqrt{1-\eta} & 0  \\
  0 & 0 & 0 & \sqrt{\eta(1-\eta)}  \\
  0 & 0 & 0 & 0  \\
  0 & 0 & 0 & 0  \\
  \end{array} \right), \nonumber
\end{eqnarray}
\begin{eqnarray}
 \hspace{-0.2cm}A_3 = E_1 \otimes E_1 = \left(
  \begin{array}{cccc}
  0 & 0 & 0 & 1-\eta  \\
  0 & 0 & 0 & 0  \\
  0 & 0 & 0 & 0  \\
  0 & 0 & 0 & 0  \\
  \end{array} \right).
\label{eq:memoryless} 
\end{eqnarray}
\normalsize

For two channel uses, a \textit{full-memory} 
amplitude damping channel was introduced in 
Ref.~\cite{yeo} and recently investigated in Ref.~\cite{Jahangir,MADC2013}
\begin{equation}
 \rho \quad \rightarrow \quad \rho'={\cal E}_1(\rho)\,=\,\sum_i B_i \,\rho\,B_i^\dag,
 \label{eq:full-memory-channel}
\end{equation}
with the Kraus operators 
\begin{equation}
{B}_0 = \left(
\begin{array}{cccc}
  1 & 0 & 0 & 0  \\
  0 & 1 & 0 & 0  \\
  0 & 0 & 1 & 0  \\
  0 & 0 & 0 & \sqrt{\eta}  \\
  \end{array} \right), \,\,
{B}_1 = \left(
\begin{array}{cccc}
  0 & 0 & 0 & \sqrt{1-\eta}  \\
  0 & 0 & 0 & 0  \\
  0 & 0 & 0 & 0  \\
  0 & 0 & 0 & 0  \\
  \end{array} \right).
\label{eq:memory-Kraus-Operators}
\end{equation}
In ${\cal E}_1$ the relaxation phenomena are fully correlated.
In other words,
when a qubit undergoes a relaxation process, the other qubit does the same.
In this way only the state $\ket{11}\equiv\ket{1}\otimes\ket{1}$  
can decay, while the other states $\ket{ij}\equiv\ket{i}\otimes\ket{j}$, 
$i,j\,\in\,\{0,1\}$, $ij \neq 11$, are not affected. 

In this paper we will focus on the partially correlated channel
${\cal E}_\mu$, defined as a convex combination of the memoryless
channel ${\cal E}_0$ and the full memory channel 
${\cal E}_1$
\begin{equation}
 \rho \quad \rightarrow \quad \rho'={\cal E}_\mu(\rho)\,=\,
(1-\mu) {\cal E}_0(\rho)+\mu\,{\cal E}_1(\rho).
\label{eq:model}
\end{equation}
Here, $\mu\,\in\,[0,1]$ is the memory parameter: the 
memoryless channel (${\cal E}_0$) is recovered when $\mu=0$, whereas for 
$\mu=1$ we obtain the ``full memory'' amplitude 
damping channel (${\cal E}_1$).
In the following, we will derive lower bounds for the
single-shot classical capacity $C_1({\cal E}_\mu)$, 
lower and upper bounds for the quantum capacity $Q({\cal E}_\mu)$,
and we will compute the entanglement-assisted classical capacity $C_E({\cal E}_\mu)$.

%\section{Channel covariance with respect to ${\cal R}_i$ and the Swap gate}
%\label{sec:channel_covariance_properties}
We will now investigate some covariance properties of the above channel, 
that will be subsequently exploited to derive the above mentioned bounds.
We define the following unitary operators:
\begin{equation} 
 {\cal R}_1=\sigma_z \otimes \openone, \quad {\cal R}_2=\openone \otimes \sigma_z, \quad
 {\cal R}_3=\sigma_z \otimes \sigma_z.
 \label{eq:R-operators}
\end{equation}

It is straightforward to demonstrate that the operators 
$A_i$ (\ref{eq:memoryless}) and $B_i$ (\ref{eq:memory-Kraus-Operators}) 
either commute or
anticommute with ${\cal R}_i$ (\ref{eq:R-operators}), namely
\begin{eqnarray} 
 &&\hspace{-1cm} A_0 {\cal R}_i={\cal R}_i A_0,\quad
                 B_0 {\cal R}_i={\cal R}_i B_0, \quad \forall\, i\,\in\{1,2,3\},\\
 &&\hspace{-1cm} {\cal R}_1 A_1\,=\, A_1  {\cal R}_1, \quad
                 {\cal R}_2 A_1\,=\, -A_1  {\cal R}_2, \quad
                 {\cal R}_3 A_1\,=\, -A_1  {\cal R}_3, \\
 &&\hspace{-1cm} {\cal R}_1 A_2\,=\ -A_2  {\cal R}_1, \quad
		 {\cal R}_2 A_2\,=\, A_2  {\cal R}_2, \quad
		 {\cal R}_3 A_2\,=\,-A_2  {\cal R}_3, \\
 &&\hspace{-1cm} {\cal R}_3 A_2\,=\, -A_3  {\cal R}_1,\quad
		 {\cal R}_2 A_3\,=\, -A_3  {\cal R}_2, \quad
		 {\cal R}_3 A_2\,=\,  A_3  {\cal R}_3, \\
 &&\hspace{-1cm} {\cal R}_1 B_1 \,=\, -B_1  {\cal R}_1,\quad
		 {\cal R}_2 B_1\,=\, -B_1  {\cal R}_2,\,\quad
                 {\cal R}_3 B_1\,=\, B_1  {\cal R}_3.
\end{eqnarray}
>From the above relations it follows that
\begin{eqnarray} 
   &&\hspace{-0.5cm}{\cal E}_0({\cal R}_1 \, \rho \, {\cal R}_1)\,=\,
                  \sum_{i=0}^3 
A_i {\cal R}_1 \, \rho \, {\cal R}_1 A_i^\dag\,=\nonumber\\
                       &&=\,   
                           {\cal R}_1 A_0 \rho A_0 {\cal R}_1 \,+\,
                           {\cal R}_1 A_1 \rho {\cal R}_1 A^\dag_1 \,+\,\nonumber\\
                          &&\hspace{1cm} +(-{\cal R}_1 A_2) \rho (-A^\dag_2 {\cal R}_1) \,+\,
                           (-{\cal R}_1 A_3) \rho (-A^\dag_3 {\cal R}_1 )\,=\nonumber\\
                       &&=\,{\cal R}_1 \Big(\sum_i A_i \, \rho \, A_i^\dag\Big) {\cal R}_1\,=
                            {\cal R}_1\, {\cal E}_0(\rho)\, {\cal R}_1,
   \label{eq:R1-inv2}
\end{eqnarray}
where we use  
\begin{eqnarray}
&& A^\dag_0=A_0, \nonumber\\
&&   {\cal R}_1 A^\dag_1=(A_1 {\cal R}_1)^\dag,=
     ({\cal R}_1 A_1)^\dag=A_1^\dag {\cal R}_1,  \nonumber\\
&&   {\cal R}_1 A^\dag_2=(A_2 {\cal R}_1)^\dag=
     (-{\cal R}_1 A_2)^\dag=-A_2^\dag {\cal R}_1, \nonumber\\
&&   {\cal R}_1 A^\dag_3=(A_3 {\cal R}_1)^\dag=
     (-{\cal R}_1 A_3)^\dag=-A_3^\dag {\cal R}_1.   \nonumber
\end{eqnarray}
In a similar way it can be shown that 
${\cal E}_0({\cal R}_2 \, \rho \, {\cal R}_2)={\cal R}_2\, {\cal E}_0(\rho)\, {\cal R}_2$
and ${\cal E}_0({\cal R}_3 \, \rho \, {\cal R}_3)={\cal R}_3\, {\cal E}_0(\rho)\, {\cal R}_3$:
the channel ${\cal E}_0$ is covariant with respect to all the operators
${\cal R}_i$.  
With a similar argument it can be proved that also the full memory channel 
${\cal E}_1$ is covariant with respect to ${\cal R}_i$ \cite{MADC2013}.
Therefore, also the channel with an arbitrary 
degree of memory is covariant with respect to ${\cal R}_i$, namely 
\begin{eqnarray}
\hspace{-1.8cm}{\cal E}_\mu({\cal R}_i\rho {\cal R}_i)=
(1-\mu)\,{\cal R}_i{\cal E}_0(\rho){\cal R}_i
  \,+\,\mu\,{\cal R}_i{\cal E}_1(\rho){\cal R}_i=
{\cal R}_i{\cal E}_\mu(\rho ){\cal R}_i.
\label{eq:Ri-cov}
\end{eqnarray}

Now we consider the action of the Swap gate~\cite{nielsen-chuang}, defined as
\begin{equation}
{\cal S}_\textrm{w}\equiv\ketbra{00}{00}\,+\,\ketbra{01}{10}\,+\,\ketbra{10}{01}\,+\,\ketbra{11}{11}.
 \label{eq:swap-gate}
\end{equation}
We notice that
\begin{eqnarray}
 &&\hspace{-1.0cm}
 {\cal S}_\textrm{w}\,A_0\, {\cal S}_\textrm{w}\,=\, A_0, \quad
 {\cal S}_\textrm{w}\,A_1\, {\cal S}_\textrm{w}\,=\, A_2, \quad
 {\cal S}_\textrm{w}\,A_2\, {\cal S}_\textrm{w}\,=\, A_1, \quad
 {\cal S}_\textrm{w}\,A_3\, {\cal S}_\textrm{w}\,=\, A_3.
\end{eqnarray}
By using ${\cal S}_\textrm{w}^\dag={\cal S}_\textrm{w}$, ${\cal S}_\textrm{w}
{\cal S}_\textrm{w}=\openone$ and the above relations,
we can easily prove that the channel ${\cal E}_0$ is covariant with respect 
to
 ${\cal S}_\textrm{w}$, namely
\begin{eqnarray} 
\hspace{-1.0cm}{\cal E}_0({\cal S}_\textrm{w} \, \rho \, {\cal S}_\textrm{w})\,=\,
%\sum_i A_i {\cal S}_\textrm{w} \, \rho \, {\cal S}_\textrm{w} A_i^\dag\,=\nonumber\\
%=\,  A_0 {\cal S}_\textrm{w} \rho {\cal S}_\textrm{w} A^\dag_0 \,+\,
%A_1 {\cal S}_\textrm{w} \rho {\cal S}_\textrm{w} A^\dag_1 \,+\,
%A_2 {\cal S}_\textrm{w} \rho {\cal S}_\textrm{w} A^\dag_2 \,+\,
%A_3 {\cal S}_\textrm{w} \rho {\cal S}_\textrm{w} A^\dag_3\,=\nonumber\\
{\cal S}_\textrm{w}\, {\cal E}_0(\rho)\, {\cal S}_\textrm{w}\;.
   \label{eq:S-inv2}
\end{eqnarray}
%where we use that 
%\begin{eqnarray}
%&& {\cal S}_\textrm{w}^\dag={\cal S}_\textrm{w}, 
%\quad {\cal S}_\textrm{w}{\cal S}_\textrm{w}=\openone \nonumber\\
%&&   A_1 {\cal S}_\textrm{w} = {\cal S}_\textrm{w} \big({\cal S}_\textrm{w} A_1 {\cal S}_\textrm{w}\big) = {\cal S}_\textrm{w} A_2 \nonumber\\
% &&  {\cal S}_\textrm{w} A_1^\dag = (A_1 {\cal S}_\textrm{w})^\dag = ({\cal S}_\textrm{w} A_2)^\dag=A_2^\dag {\cal S}_\textrm{w}  \nonumber\\
%&&   A_2 {\cal S}_\textrm{w} = {\cal S}_\textrm{w} \big({\cal S}_\textrm{w} A_2 {\cal S}_\textrm{w}\big) = {\cal S}_\textrm{w} A_1  \nonumber\\
% &&  {\cal S}_\textrm{w} A_2^\dag = (A_2 {\cal S}_\textrm{w})^\dag = ({\cal S}_\textrm{w} A_1)^\dag=A_1^\dag {\cal S}_\textrm{w}  \nonumber
%\end{eqnarray}
It is straightforward to demonstrate that ${\cal S}_\textrm{w}$ commutes 
with the Kraus operators $B_0$ and $B_1$ (\ref{eq:memory-Kraus-Operators}). 
Therefore the channel ${\cal E}_1$
is covariant with respect to ${\cal S}_\textrm{w}$.
Since both the channels ${\cal E}_0$ and ${\cal E}_1$ are covariant with 
respect to ${\cal S}_\textrm{w}$, the channel ${\cal E}_\mu$ is also
covariant under the action of ${\cal S}_\textrm{w}$.

\section{Classical capacity}
\label{sec:classical-capacity}

In this section we will study the performance of the channel
to transmit classical information, quantified by the classical capacity
$C$, that measures the maximum amount of classical information that can be 
reliably transmitted down the channel per channel use. 
More specifically, we address the problem
of computing the single shot capacity $C_1$~\cite{nielsen-chuang} 
of the partially correlated channel ${\cal E}_\mu$, that is achieved
by maximizing the so called Holevo quantity 
$\chi$~\cite{nielsen-chuang,benenti-casati-strini,holevo73,
schumacher-westmoreland,holevo98,hausladen}
with respect to one use of the channel ${\cal E}_\mu$ as follows:
\begin{eqnarray}
  &&C_1({\cal E}_\mu)=\max_{\{p_\alpha,\rho_\alpha\}}{\chi\big({\cal E}\mu,\{p_\alpha,\rho_\alpha\}\big)}.
    \label{eq:C1} 
\end{eqnarray}
In the above expression $\{p_\alpha, \rho_\alpha\}$ is a quantum source, 
described by the 
density operator $\rho=\sum_\alpha p_\alpha \rho_\alpha$ and 
the Holevo quantity is defined as
\begin{eqnarray}
  &&\hspace{-1cm}\chi\big({\cal E}\mu,\{p_\alpha,\rho_\alpha\}\big) 
      \equiv S({\cal E}\mu(\rho)) \, - \, \sum_\alpha p_\alpha S({\cal E}\mu(\rho_\alpha)),
    \label{eq:chi}
\end{eqnarray}
where $S(\rho)=-{\rm Tr}(\rho\log_2 \rho)$ 
is the von Neumann entropy.
Without loss of generality, in the following we will restrict to ensembles of 
pure states $\{p_k,\ket{\psi_k}\}$, since any ensemble of mixed states can be 
described by an ensemble of pure states with same density operator, and
whose Holevo quantity (\ref{eq:chi}) is at least as 
large~\cite{schumacher-westmoreland}.
The above expressions then become 
\begin{eqnarray}
  &&\hspace{-0.5cm} C_1({\cal E}\mu)=\max_{\{p_k,\ket{\psi_k}\}}{\chi\big({\cal E}\mu,\{p_k,\ket{\psi_k}\}\big)},
    \label{eq:C1-b} \\
  &&\hspace{-0.5cm}\chi\big({\cal E}\mu,\{p_k,\ket{\psi_k}\}\big)\,= \nonumber \\
  &&\hspace{0.5cm}=\,S({\cal E}\mu(\rho)) \, - \, \sum_k p_k S({\cal E}\mu(\ketbra{\psi_k}{\psi_k})),
    \label{eq:chi-pure}
\end{eqnarray}
where now $\rho=\sum_k\, p_k\ketbra{\psi_k}{\psi_k}$.
The optimization of $C_1$ was performed for the amplitude damping channel
with full memory ($\mu=1$) in Ref. \cite{MADC2013}. The case of partial memory
is harder to treat, so in the following we will derive lower bounds 
on $C_1$, by exploiting the channel covariance properties discussed above
and employing specific ensembles.

\subsection{Form of optimal ensembles}
\label{sec:max-chi}

We derive here a general form of the ensemble that optimizes the Holevo
quantity, by exploiting the covariance properties discussed in the previous 
section.
First we take advantage of the covariance property of the channel
${\cal E}_\mu$ with respect to ${\cal R}_i$~(\ref{eq:R-operators}).
Given a generic ensemble $\{p_k,\ket{\psi_k}\}$, we consider a new ensemble  
by replacing 
each state $\ket{\psi_k}$ in $\{p_k,\ket{\psi_k}\}$ by the set
\begin{eqnarray}
\{\ket{\psi_k},\, {\cal R}_1\ket{\psi_k},\,{\cal R}_2\ket{\psi_k},\,{\cal R}_3\ket{\psi_k}\},
\nonumber
\end{eqnarray}
each state occurring with probability $\tilde{p}_k=p_k/4$. We refer to
this new ensemble as $\{\tilde{p}_k,\ket{\tilde{\psi}_k}\}$, and call 
$\tilde{\rho}=\sum_k \tilde{p}_k \ketbra{\tilde{\psi}_k}{\tilde{\psi}_k}$ 
the associated density operator
%We observe that
\begin{eqnarray}
   %  &&\hspace{-1.0cm} \rho = \sum_k p_k \ketbra{\psi_k}{\psi_k} \rightarrow \nonumber\\
     &&\hspace{-0.5cm} \tilde{\rho}\,=\,\sum_k \frac{p_k}{4} \Big(\ketbra{\psi_k}{\psi_k}\,+\,
                             \sum_{i=1}^3{\cal R}_i\ketbra{\psi_k}{\psi_k}{\cal R}_i\Big)\,=\nonumber\\
     &&\hspace{-0.1cm}       =\frac{1}{4}\Big(\rho \,+\,\sum_{i=1}^3{\cal R}_i\rho{\cal R}_i\Big).
    \label{eq:rho-tilde-a}
\end{eqnarray}
%and 
It can be verified that $\tilde{\rho}$ %(\ref{eq:rho-tilde-a}) 
has the same diagonal elements of $\rho$,
while the off-diagonal entries are all vanishing.
We now show that
\begin{equation}
   \chi\big({\cal E}\mu,\{\tilde{p}_k,\ket{\tilde{\psi}_k}\}\big) \ge 
\chi\big({\cal E}\mu,\{p_k,\ket{\psi_k}\}\big).
\label{eq_obj1}
\end{equation}
To this end we first notice that
% which describes the new ensemble.
%By using to the identities (\ref{eq:Ri-cov}), it turns out that
\begin{eqnarray}
&&\hspace{-0.5cm}
%S\big({\cal E}\mu(\ketbra{\tilde{\psi}_k}{\tilde{\psi}_k})\big)\,=\,
S\big({\cal E}\mu({\cal R}_i \ketbra{\psi_k}{\psi_k} {\cal R}_i)\big) \,=\,
S\big({\cal R}_i\,{\cal E}\mu(\ketbra{\psi_k}{\psi_k})\,{\cal R}_i\big)
 \nonumber\\
&&\hspace{2.7cm}=\,
S\big({\cal E}\mu(\ketbra{\psi_k}{\psi_k})\big),
\label{equalities-a}
\end{eqnarray}
where we used Eqs.~(\ref{eq:Ri-cov}) and the fact that 
a unitary operation does not change the von Neumann entropy. 
Therefore, by replacing the old ensemble with the new one, 
the second term in the Holevo quantity 
(\ref{eq:chi-pure}) does not change, namely
\begin{eqnarray}
      && \hspace{-2.5cm}\sum_k \tilde{p}_k S({\cal E}\mu(\ketbra{\tilde{\psi}_k}{\tilde{\psi}_k}))\,
       =\,4 \sum_k \frac{p_k}{4} S({\cal E}\mu(\ketbra{\psi_k}{\psi_k}))\,=\,
      \sum_k p_k S({\cal E}\mu(\ketbra{\psi_k}{\psi_k})).
    \label{eq:chi-second-term-a}
\end{eqnarray}
For the output entropy related to $\tilde{\rho}$ we have
\begin{eqnarray}
     &&\hspace{-1.5cm} S({\cal E}\mu(\tilde{\rho}))\,=\,
            S\Big({\cal E}\mu\Big(\frac{1}{4}\rho \,+\,\frac{1}{4}\sum_{i=1}^3{\cal R}_i\rho{\cal R}_i \Big)\Big)
                \,=\,
            S\Big(\frac{1}{4}{\cal E}\mu\big(\rho\big) \,+\,
                 \frac{1}{4}\sum_{i=1}^3{\cal E}\mu\big({\cal R}_i\rho{\cal R}_i\big)\Big) \,  \nonumber\\
     &&\hspace{0.2cm} \ge 
            \frac{1}{4}S\big({\cal E}\mu\big(\rho\big)\big) \,+\,
               \frac{1}{4}\sum_{i=1}^3 S\big({\cal E}\mu\big({\cal R}_i\rho{\cal R}_i\big)\big) 
      \,=\, S\big({\cal E}\mu\big(\rho\big)\big),
     \label{eq:chi-first-term-a}
\end{eqnarray}
where we used the linearity of %the quantum operation
 ${\cal E}_\mu$, the concavity of the von Neumann 
entropy~\cite{nielsen-chuang}, and  Eq.~(\ref{equalities-a}).
%Eqs.~ (\ref{eq:chi-second-term-a}) 
%and (\ref{eq:chi-first-term-a}) imply that
%\begin{equation}
%   \chi\big({\cal E}_m,\{\tilde{p}_k,\ket{\tilde{\psi}_k}\}\big) \ge \chi\big({\cal E}_m,\{p_k,\ket{\psi_k}\}\big).
%\end{equation}
Relations (\ref{eq:chi-second-term-a}) 
and (\ref{eq:chi-first-term-a}) then prove the inequality~(\ref{eq_obj1}).
%and we can summarize 
%the conclusions as follows:
%As a conclusion of the above discussion we can state that 
In other words, for an arbitrary ensemble 
of pure states we can always find another ensemble,
whose density matrix has the same diagonal elements as the initial ensemble 
and vanishing off-diagonal entries, and whose Holevo quantity is at least as 
large.

We will now take advantage of the covariance of the channel
${\cal E}_\mu$ with respect to the swap gate 
${\cal S}_\textrm{w}$~(\ref{eq:swap-gate}).
Given a quantum ensemble $\{\tilde{p}_k,\ket{\tilde{\psi}_k}\}$, with
$\tilde{\rho}=\sum_k \tilde{p}_k \ketbra{\tilde{\psi}_k}{\tilde{\psi}_k}=$
 $\,$ $\textrm{diag}\{\alpha,\beta,\gamma,\delta\}$, 
we construct a new ensemble $\{\bar{p}_k,\ket{\bar{\psi}_k}\}$ by replacing 
each state $\ket{\tilde{\psi}_k}$ in $\{\tilde{p}_k,\ket{\tilde{\psi}_k}\}$ 
with the following couple of states
\begin{eqnarray}
\{\ket{\tilde{\psi}_k},\, {\cal S}_\textrm{w} \ket{\tilde{\psi}_k}\},
\nonumber
\end{eqnarray}
each state occurring with probability $\bar{p}_k=\tilde{p}_k/2$. 
We refer to this new ensemble as $\{\bar{p}_k,\ket{\bar{\psi}_k}\}$, and 
call 
$\bar{\rho}=\sum_k \bar{p}_k \ketbra{\bar{\psi}_k}{\bar{\psi}_k}$ the density
operator which describes it.
We now show that
\begin{equation}
   \chi\big({\cal E}\mu,\{\bar{p}_k,\ket{\bar{\psi}_k}\}\big) \ge \chi\big({\cal E}\mu,\{\tilde{p}_k,\ket{\tilde{\psi}_k}\}\big).
\label{eq_obj2}
\end{equation}
In order to do this, we first exploit the covariance property of the channel 
with respect to ${\cal S}_\textrm{w}$~(\ref{eq:S-inv2}), which leads to
\begin{eqnarray}
&&\hspace{-1cm}
%S\big({\cal E}_\mu(\ketbra{\bar{\psi}_k}{\bar{\psi}_k})\big)\,=\,
S\big({\cal E}_\mu({\cal S}_\textrm{w} \ketbra{\tilde{\psi}_k}{\tilde{\psi}_k} {\cal S}_\textrm{w})\big) 
\,=\,S\big({\cal S}_\textrm{w}\,{\cal E}_\mu(\ketbra{\tilde{\psi}_k}{\tilde{\psi}_k})\,{\cal S}_\textrm{w}\big)\,
 \nonumber\\
&&\hspace{2.1cm}=\,
S\big({\cal E}_\mu(\ketbra{\tilde{\psi}_k}{\tilde{\psi}_k})\big).
\label{equalities-a2}
\end{eqnarray} 
Therefore, by replacing the old ensemble by the new one, the second term in the Holevo quantity 
(\ref{eq:chi-pure}) does not change, namely
\begin{eqnarray}
      &&\hspace{-2cm}\sum_k \bar{p}_k S({\cal E}_\mu(\ketbra{\bar{\psi}_k}{\bar{\psi}_k}))
       =2 \sum_k \frac{\tilde{p}_k}{2} S({\cal E}_\mu(\ketbra{\tilde{\psi}_k}{\tilde{\psi}_k}))\,=\,
       \sum_k \tilde{p}_k S({\cal E}_\mu(\ketbra{\tilde{\psi}_k}{\tilde{\psi}_k})).
    \label{eq:chi-second-term-a2}
\end{eqnarray}
Let us now consider the changes in the first term of the Holevo quantity 
(\ref{eq:chi-pure}). First we consider the relation between
$\bar{\rho}$ and $\tilde{\rho}$, namely
\begin{eqnarray}
     &&\hspace{-1.5cm} \tilde{\rho} = \sum_k \tilde{p}_k \ketbra{\tilde{\psi}_k}{\tilde{\psi}_k} \rightarrow \nonumber\\
     &&\hspace{-1.0cm}  \bar{\rho}\,=\,\sum_k \frac{\tilde{p}_k}{2} \Big(\ketbra{\tilde{\psi}_k}{\tilde{\psi}_k}\,+\,
                             {\cal S}_\textrm{w}\ketbra{\tilde{\psi}_k}{\tilde{\psi}_k}{\cal S}_\textrm{w}\Big) 
     =\frac{1}{2}\Big(\tilde{\rho} \,+\,{\cal S}_\textrm{w}\tilde{\rho}{\cal S}_\textrm{w}\Big)\,=\nonumber\\
   &&= \left(
\begin{array}{cccc}
  \alpha & 0 & 0 & 0  \\
  0 & \frac{\beta+\gamma}{2} & 0 & 0  \\
  0 & 0 & \frac{\beta+\gamma}{2} & 0  \\
  0 & 0 & 0 & \delta  \\
  \end{array} \right).
    \label{eq:rho-tilde-a2}
\end{eqnarray}
We have that
\begin{eqnarray}
     &&\hspace{-0.8cm} S({\cal E}_\mu(\bar{\rho}))\,=\,
            S\Big({\cal E}_\mu\Big(\frac{1}{2}\tilde{\rho} \,+
               \,\frac{1}{2}{\cal S}_\textrm{w}\tilde{\rho}{\cal S}_\textrm{w}\Big)\Big) \,=\,
            S\Big(\frac{1}{2}{\cal E}_\mu\big(\tilde{\rho}\big) \,+
               \,\frac{1}{2}{\cal E}_\mu\big({\cal S}_\textrm{w}\tilde{\rho}{\cal S}_\textrm{w}\big)\Big) \nonumber\\
     &&\hspace{+0.8cm} \ge 
            \frac{1}{2}S\big({\cal E}_\mu\big(\tilde{\rho}\big)\big) \,+
          \,\frac{1}{2}S\big({\cal E}_\mu\big({\cal S}_\textrm{w}\tilde{\rho}{\cal S}_\textrm{w}\big)\big) 
     \,= \, S\big({\cal E}_\mu\big(\tilde{\rho}\big)\big).
     \label{eq:chi-first-term-a2}
\end{eqnarray}
Relations (\ref{eq:chi-second-term-a2}) and (\ref{eq:chi-first-term-a2})
then prove inequality (\ref{eq_obj2}).
We can summarize the above argument as follows: for any quantum ensemble 
of pure states we can find another ensemble, whose density
matrix has the same diagonal as the original one, with zero off-diagonal 
entries, with equal populations for the states $\ket{01}$ and $\ket{10}$,
and whose Holevo quantity is at least as large.
In the following we will consider such kind of ensembles, which we will 
indicate by 
$\{p_k,\ket{\psi_k}\}$.  
%\begin{equation}
%   \chi\big({\cal E}_m,\{\tilde{p}_k,\ket{\tilde{\psi}_k}\}\big) \ge \chi\big({\cal E}_m,\{p_k,\ket{\psi_k}\}\big)
%\end{equation}
%So, in the following, we assume to work with ensembles whose density matrix is diagonal.
A generic input state $\ket{\psi_k}$ in these ensembles has the form
\begin{equation}
    \ket{\psi_k}\,=\,a_k \ket{00}\,+\,b_k\ket{01}\,+\,c_k\ket{10}\,+\,d_k\ket{11},
  \label{eq:generic-input-state2}
\end{equation}
where the coefficients $a_k,\,b_k,\,c_k,\,d_k\ \in \mathbb{C}$ and satisfy
the normalization condition $|a_k|^2+|b_k|^2+|c_k|^2+|d_k|^2=1$. 
The corresponding density matrix is given by
\begin{equation}
\rho = \left(
\begin{array}{cccc}
  \alpha & 0 & 0 & 0  \\
  0 & \beta & 0 & 0  \\
  0 & 0 & {\beta} & 0  \\
  0 & 0 & 0 & \delta  \\
  \end{array} \right),
\label{eq:diagonal-density-operator2} 
\end{equation}
where
\begin{eqnarray}
&&\hspace{-2.4cm}\alpha=\sum_k p_k |a_k|^2, \quad {\beta}=\sum_k p_k |b_k|^2=\sum_k p_k |c_k|^2, \quad
 \delta=\sum_k p_k |d_k|^2\,=\,1\,-\,\alpha\,-\,2{\beta}.
\label{eq:populations2} 
\end{eqnarray}

\subsection{Lower bounds for $C_1({\cal E}_\mu)$}

Computing the $C_1$ capacity for the channel ${\cal E}_\mu$ is a very hard 
task
since one should perform the following maximization:
\begin{equation}
C_1({\cal E}_\mu)=\max_{\{p_k,\ket{\psi_k}\}} \chi\big({\cal E}_\mu,
\{p_k,\ket{\psi_k}\}\big)
\label{eq:C1Emu}
\end{equation}
over all quantum ensembles $\{p_k,\ket{\psi_k}\}$ of the 
form (\ref{eq:generic-input-state2}, \ref{eq:diagonal-density-operator2}).
We will derive here some lower bounds for $C_1({\cal E}_\mu)$ by optimizing 
the Holevo quantity of ${\cal E}_\mu$ with respect to
some specific ensembles of the above mentioned form.
We will consider two types of such ensembles.

The first ensemble, which we call ${\cal G}_1$, is given by the following 
eight states:
\begin{eqnarray}
    && \ket{\psi},\,\,{\cal R}_i\ket{\psi},\,\,
       {\cal S}_\textrm{w}\ket{\psi}, \,\,{\cal R}_i{\cal S}_\textrm{w}\ket{\psi},
     \quad i \in \{1,2,3\}  \label{eq:statesG1},\\
  && \textrm{where}\quad \ket{\psi}\,=\,a \ket{00}\,+\,b\ket{01}\,+\,c\ket{10}\,+\,d\ket{11},
  \label{eq:generic-input-stateG1}
\end{eqnarray}
each state occurring with the same probability $p=1/8$. 
Here $a, b, c, d$ are complex numbers. It is straightforward to show that
the resulting density matrix $\rho$ has the form 
(\ref{eq:diagonal-density-operator2})
%$\rho=\textrm{diag}\{\alpha, \tilde{\beta},\tilde{\beta},{\delta}\}$ 
where
\begin{eqnarray}
&&\alpha=|a|^2, \quad {\beta}=\frac{|b|^2+|c|^2}{2}, \quad \delta=|d|^2.
\label{eq:populationsG1} 
\end{eqnarray}
The Holevo quantity relative to the ensemble ${\cal G}_1$ is
\begin{equation}
\chi\big({\cal E}_\mu,{\cal G}_1)\,=\,S\big({\cal E}_\mu(\rho)\big)-
S\big({\cal E}_\mu(\ketbra{\psi}{\psi})\big),
\end{equation}
since by construction all the states in the ensemble
  (\ref{eq:generic-input-stateG1}) have the same output entropy, due to
the covariance properties of ${\cal E}_\mu$ with respect to ${\cal R}_i$ 
(\ref{eq:Ri-cov}) and ${\cal S}_\textrm{w}$ (\ref{eq:S-inv2}) exploited above.
Therefore a lower bound for the classical capacity of the channel 
${\cal E}_\mu$ can be derived from
\begin{equation}
\chi_{lwb_{{\cal G}_1}}(\eta,\mu)=\max_{a,b,c,d}\chi\big({\cal E}_\mu,{\cal G}_1).
\label{eq:max_chi_G1}
\end{equation}
Without loss of generality we set
\begin{eqnarray}
&& a=\overline{a},\quad b=\overline{b}e^{i\varphi_1}, \quad c=\overline{c}e^{i\varphi_2},
\quad d=\sqrt{1-\overline{a}^2-\overline{b}^2-\overline{c}^2}\,e^{i\varphi_3},\, \quad \,\nonumber\\
&& \textrm{and} \quad \overline{a},\,\,\overline{b},\,\,\overline{c},\,\,\varphi_1,\,\,\varphi_2,\,\,\varphi_3
\in \mathbb{R}.
\label{eq:coefficientsforG1}
\end{eqnarray}
The maximization (\ref{eq:max_chi_G1}) can be recast as
\begin{equation}
\chi_{lwb_{{\cal G}_1}}(\eta,\mu)=\max_{\overline{a},\,\overline{b},\,\overline{c},\,
\varphi_1,\,\varphi_2,\,\varphi_3}\chi\big({\cal E}_\mu,{\cal G}_1),
\label{eq:max_chi_G1-2}
\end{equation}
with the following constraints:
\begin{eqnarray}
\overline{a},\,\,\overline{b},\,\,\overline{c} \,\,\in \,\, [0,1], \quad
\overline{a}^2+\overline{b}^2+\overline{c}^2 \, \le \,1, \quad
 \varphi_i \,\, \in \,\, [0,2\pi[. 
\end{eqnarray}
The reason for investigating such a lower bound is that the set ${\cal G}_1$ 
contains the ensemble which allows to achieve the product state 
capacity $2\,C_{madc,1}$~\cite{nielsen-chuang}
for two uses of the memoryless amplitude damping channel~\cite{giovannetti}.
In the case of memoryless channel ($\mu=0$) the lower bound
(\ref{eq:max_chi_G1}) will be at least equal to $2\,C_{madc,1}$.

The second quantum ensemble we consider is of the kind
\begin{eqnarray}
    \label{eq:ensemble-lowerbound}
   \{{p}_k,\ket{{\psi}_k}\}\,=\, \{{p}_{\varphi k},\ket{{\varphi}_k}\} \,\cup \, 
                                  \{{p}_{\phi, k},\ket{{\phi}_k}\}, 
 \label{eq:ensemble-G2}
\end{eqnarray}
where
\begin{eqnarray}
&&\hspace{-0.5cm}
   \left\{
   \begin{array}{ll}
    {p}_{\varphi \pm}=\beta, & \ket{\varphi_{+}}\,=\,
                             \cos\theta_1\ket{01}\,+\, e^{i\varphi_1}\sin\theta_1\ket{10},\\
                             & \ket{\varphi_{-}}\,=\,-
                             \sin\theta_1\ket{01}\,+\, e^{i\varphi_1}\cos\theta_1\ket{10},\\
    \quad & \quad \\
    {p}_{\phi \pm}= \frac{1-2\beta}{2},  &
               \ket{\phi_{\pm}}=\cos\theta_2\ket{00}\pm e^{i\varphi_2}\sin\theta_2\ket{11}.
               \end{array} \right.
               \label{eq:states-G2}
\end{eqnarray}
We call this ensemble ${\cal G}_2$.
The corresponding density operator is 
\begin{equation}
\rho = \left(
\begin{array}{cccc}
  (1-2\beta)\cos^2\theta_2 & 0 & 0 & 0  \\
  0 & {\beta} & 0 & 0  \\
  0 & 0 & {\beta} & 0  \\
  0 & 0 & 0 & (1-2\beta)\sin^2\theta_2  \\
  \end{array} \right),
\label{eq:diagonal-density-operatorG2} 
\end{equation}
which is of the form (\ref{eq:diagonal-density-operator2}).
The Holevo quantity relative to the ensemble ${\cal G}_2$
is given by
\begin{eqnarray}
&&\hspace{0cm}\chi\big({\cal E}_\mu,{\cal G}_2)\,=\,S\big({\cal E}_\mu(\rho)\big)\,+\nonumber\\
&&\hspace{1cm}-2\beta S\big({\cal E}_\mu(\ketbra{\varphi_\pm}{\varphi_\pm})\big)\,-\,
(1-2\beta) S\big({\cal E}_\mu(\ketbra{\phi_\pm}{\phi_\pm})\big),
\end{eqnarray}
since the states $\ket{\varphi_\pm}$ have the same 
output entropy, and the same for $\ket{\phi_\pm}$.
It is possible to show that any state
in the subspace spanned by $\{\ket{01},\ket{10}\}$ has the same output 
entropy, 
which only depends on the channel transmissivity $\eta$ and the 
channel degree of memory $\mu$. In other words, the entropy 
$S\big({\cal E}_\mu(\ketbra{\varphi_\pm}{\varphi_\pm})\big)$
does not depend on $\theta_1,\,\varphi_1$. Moreover, the output entropy
$S\big({\cal E}_\mu(\ketbra{\phi_\pm}{\phi_\pm})\big)$ does not depend on 
$\varphi_2$. Therefore the lower bound (\ref{eq:max_chi_G1-2})
for the classical capacity of the channel ${\cal E}_\mu$ can be computed as
\begin{equation}
\chi_{lwb_{{\cal G}_2}}(\eta,\mu)=\max_{\beta,\,\theta_2}\chi\big({\cal E}_\mu,{\cal G}_2).
\label{eq:max_chi_G2}
\end{equation}
The reason to investigate this lower bound is that the ensemble ${\cal G}_2$
contains the ensemble which allows to achieve the $C_1$ classical capacity of 
the full memory channel ${\cal E}_1$~\cite{MADC2013}, since
$\chi_{lwb_{{\cal G}_2}}(\eta,1)$ coincides with $C_1({\cal E}_1)$.

The two lower bounds (\ref{eq:max_chi_G1-2}) and (\ref{eq:max_chi_G2}) were 
computed numerically. In the following subsection we report the corresponding results.

\subsection{Numerical results}

\begin{figure}[t!]
  \begin{center}
  \includegraphics[width=9cm]{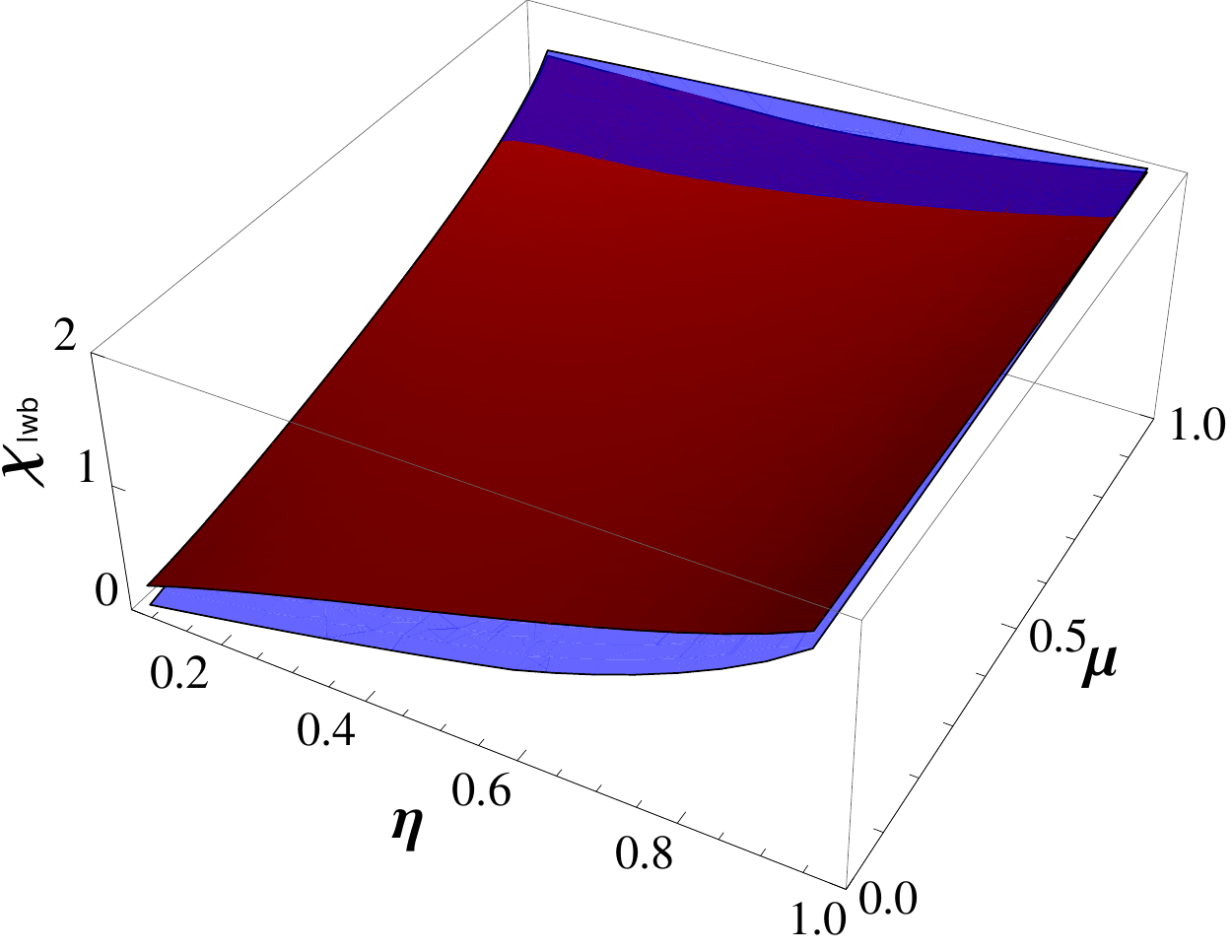}
  \end{center}
  \caption{(color online) Lower bounds $\chi_{lwb_{{\cal G}_1}}(\eta,\mu)$ 
  (\ref{eq:max_chi_G1-2}) (red surface)
  and $\chi_{lwb_{{\cal G}_2}}(\eta,\mu)$ (\ref{eq:max_chi_G2}) 
  (blue surface).
  For small $\mu$, $\chi_{lwb_{{\cal G}_1}}>\chi_{lwb_{{\cal G}_2}}$}
  \label{fig:C1lwb}
\end{figure}

In Fig.~\ref{fig:C1lwb} we plot 
the numerical results for the maximization in Eqs. 
(\ref{eq:max_chi_G1-2}) and (\ref{eq:max_chi_G2}).
As we can see, for not too high values of the memory degree ($\mu<0.8$) 
we have that $\chi_{lwb_{{\cal G}_1}}>\chi_{lwb_{{\cal G}_2}}$:
the ensemble ${\cal G}_1$ allows to achieve better performance
with respect to the ensemble ${\cal G}_2$ in transmitting
classical information across the channel ${\cal E}_\mu$. 
Instead, as expected,
the ensemble ${\cal G}_2$ is better than ${\cal G}_1$ for higher values 
of the memory degree because it is the ensemble that maximises the performance
of the full memory channel.
Moreover, since both $\chi_{lwb_{{\cal G}_1}}$ and $\chi_{lwb_{{\cal G}_2}}$ 
are increasing functions of $\mu$,
our results show that memory increases the channel aptitude
to transmit classical information. 
It is worth discussing the particular case $\eta=0$.
In the memoryless case for $\eta=0$ there is no classical information 
transmission, 
since the output state is always $\ket{00}$ for any input.
On the other hand, we can see from Fig.~\ref{fig:C1lwb} 
that any finite degree of memory allows for
information transmission also in this limiting case. 

\begin{figure}[t!]
  \begin{center}
  \includegraphics[width=10.cm]{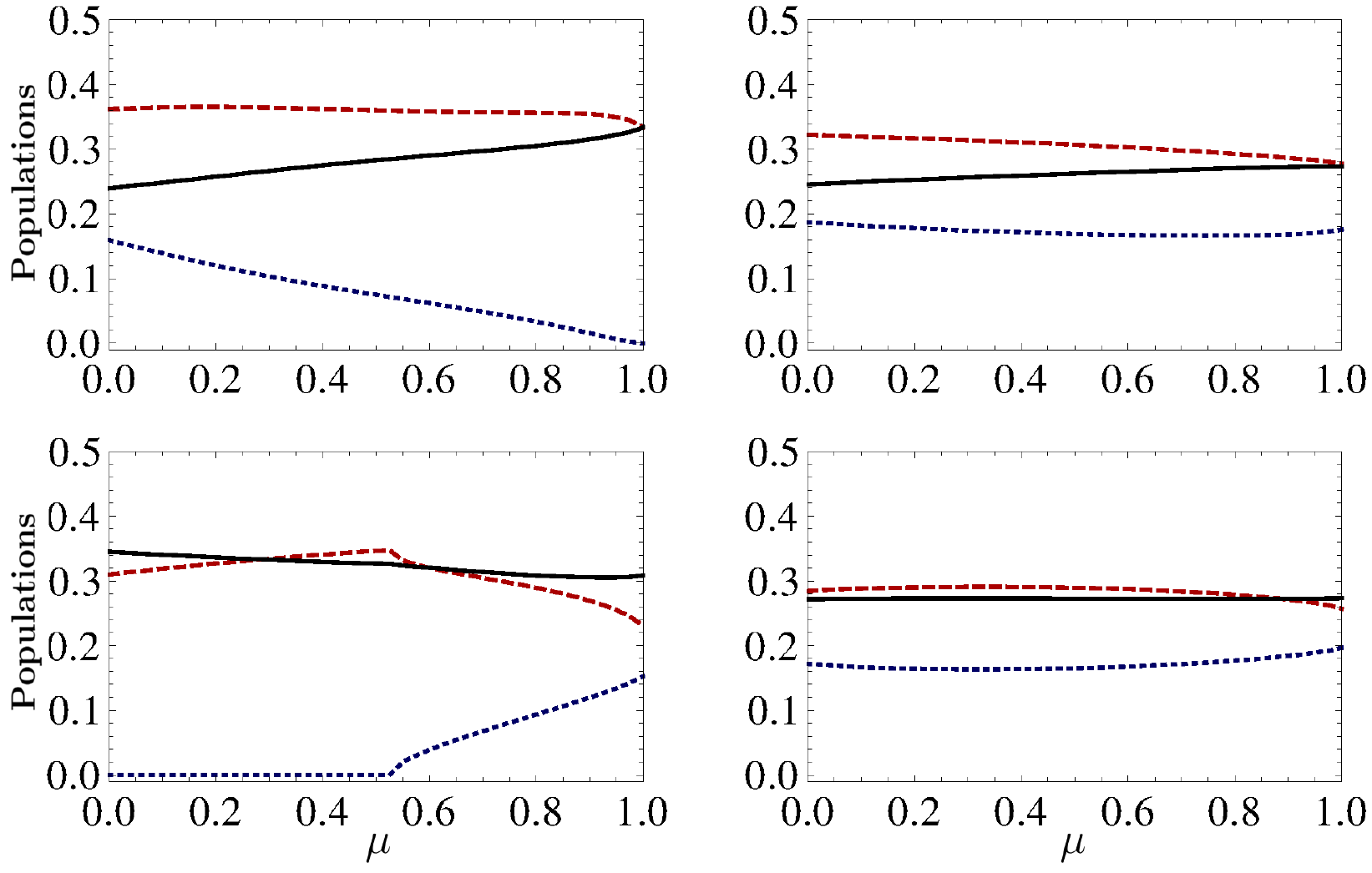}
  \end{center}
  \caption{(color online) Populations which maximize the Holevo quantity for 
          the ensemble ${\cal G}_1$ (top plots) and
          for ${\cal G}_2$ (bottom plots),
          at channel transmissivity $\eta=0.3$ (left) and $\eta=0.8$ (right),
          as functions of the memory degree $\mu$.
          For the ensemble ${\cal G}_1$: $\alpha$ (red long-dashed curve), 
           $\beta=\gamma$ (black full curve),  and
           $\delta$ (blue dashed curve).  
          For the ensemble ${\cal G}_2$: $\alpha=(1-2\beta)\cos^2\theta_2$ 
          (red long-dashed curve), $\beta$ (black full curve),
           $\delta=(1-2\beta)\sin^2\theta_2$ (blue dashed curve).}
%           (\ref{eq:diagonal-density-operatorG2}) of the ensemble ${\cal G}_2$
%           (\ref{eq:ensemble-G2})-(\ref{eq:states-G2})
%           which maximize the Holevo quantity for ${\cal E}_\mu$ (\ref{eq:max_chi_G2}),
  \label{fig:Populations}
\end{figure}

In Fig. \ref{fig:Populations} (top) we plot
the populations (\ref{eq:populationsG1}) of the ensemble ${\cal G}_1$ 
(\ref{eq:statesG1})-(\ref{eq:generic-input-stateG1})
which solves the optimization problem (\ref{eq:max_chi_G1-2}).
The populations are plotted as functions of the memory degree $\mu$, for 
two values of the channel transmissivity: $\eta=0.3$ (left plot) and
$\eta=0.8$ (right plot). 
>From the numerical optimization it turns out that
states of the optimal ensemble 
(\ref{eq:statesG1})-(\ref{eq:generic-input-stateG1}) exhibit the same
weights for the components $\ket{01}$ and $\ket{10}$ ($|b|^2=|c|^2$).
Note also that for low values of the channel transmissivity 
($\eta=0.3$ in the left plot) and 
for $\mu\approx1$,
the states (\ref{eq:generic-input-stateG1}) have vanishing components 
along $\ket{11}$; indeed for small values of the transmissivity, when 
$\mu$ approaches 1, the subspace spanned by 
$\{\ket{00},\ket{01},\ket{10}\}$ becomes noiseless, and it is not convenient 
to use the state $\ket{11}$ to encode information. 
In this last
case the bound $\chi_{lwb_{{\cal G}_1}}$ is close to $\log_23$.
It is worth noting that from numerical analysis it turns out that the 
maximum (\ref{eq:max_chi_G1-2}) is also reached for 
$\varphi_1=\varphi_2=\varphi_3=0$ (which means that
the maximum of the Holevo quantity is reached for real coefficients
$a=\bar{a},b=\bar{b},c=\bar{c}$). % (\ref{eq:coefficientsforG1}).  

In Fig.~\ref{fig:Populations} (bottom panels) we plot
the populations of the ensemble ${\cal G}_2$ 
(\ref{eq:ensemble-G2})-(\ref{eq:states-G2})
which solve the optimization (\ref{eq:max_chi_G2}). 
It is interesting to notice that for low values of the channel 
transmissivity ($\eta=0.3$ in the figure), 
the state $\ket{11}$ %(\ref{eq:diagonal-density-operatorG2}) 
is not populated for low values of the memory degree, and it is 
``activated" for a large enough 
degree of memory. In other words, for $\eta\lesssim 0.6$, we can identify a
threshold value $\mu_{th}(\eta)$ below which $\ket{11}$ 
is not populated; it turns out that 
the smaller is $\eta$, the greater is $\mu_{th}$.

We investigate the amount of entanglement required for the transmission
of classical information by considering the average entanglement of the  
quantum ensemble $\{p_k,\ket{\psi_k}\}$ employed, defined as
\begin{equation}
 E_{\{p_k,\ket{\psi_k}\}}=\sum_k p_k E(\ket{\psi_k}),
\label{eq:average entanglement}
\end{equation}
where $E(\ket{\psi_k})$ is the entropy of 
entanglement~\cite{HorodeckiReview} 
of the bipartite pure state $\ket{\psi_k}$.
The entanglement related to
the ensemble ${\cal G}_1$ is simply the entanglement of the state 
$\ket{\psi}$ in (\ref{eq:generic-input-stateG1})
\begin{equation}
 E_{{\cal G}_1}=E(\ket{\psi}),
\label{eq:G1average entanglement}
\end{equation}
since all the states (\ref{eq:statesG1})
have the same entanglement (${\cal R}_i$ are
local unitary operations, and it is simple to verify that ${\cal S}_\textrm{w}$ does not change the entanglement
of the pure state $\ket{\psi}$).
Instead, the average entanglement of the ensemble 
 ${\cal G}_2$ (\ref{eq:ensemble-G2})-(\ref{eq:states-G2})
is given by
\begin{equation}
E_{{\cal G}_2}\,=\,(1-2\beta)E(\ket{\phi_\pm}),
\label{eq:entaglementG2}
\end{equation}
since one can always choose separable states inside the subspace spanned 
by $\{\ket{01},\ket{10}\}$ and therefore the states 
$\{\ket{\varphi_\pm}\}$ in the ensemble (\ref{eq:states-G2})
do not contribute to the average entanglement, and the probability of using a 
state $\ket{\phi_\pm}$ (\ref{eq:states-G2}) is
$1-2\beta$ (the states $\ket{\phi_\pm}$ have the same entanglement). 

\begin{figure}[h!]
  \begin{center}
  \includegraphics[width=10.0cm]{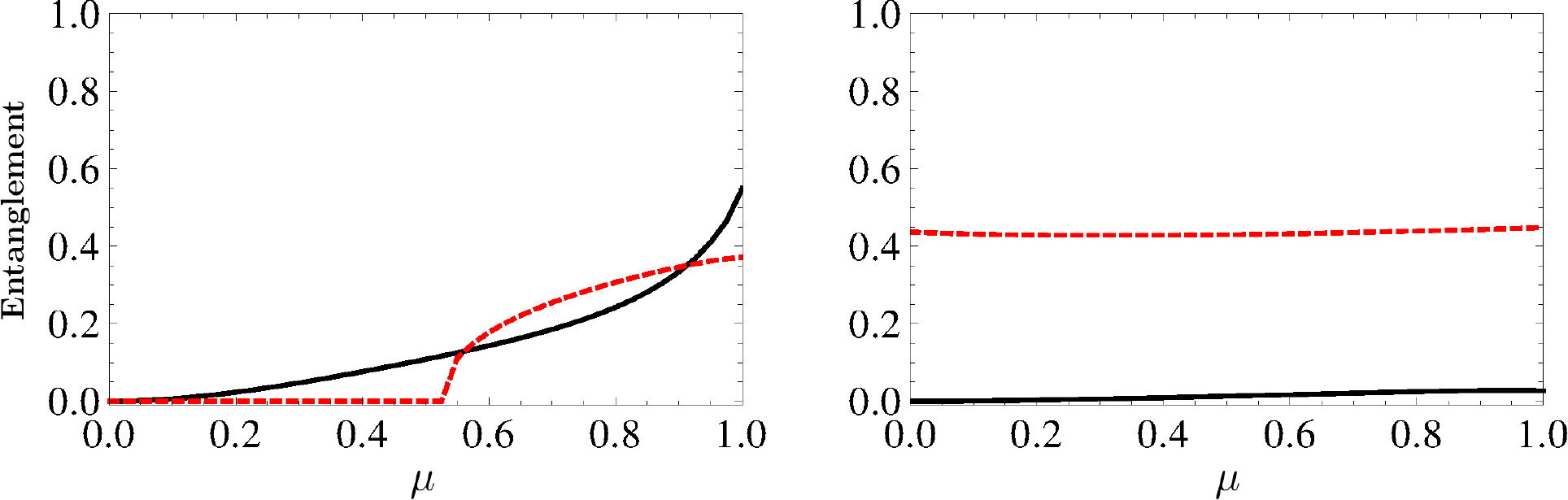}
  \end{center}
  \caption{(color online) Average entanglement of the ensembles ${\cal G}_1$
           (black full curve) and ${\cal G}_2$ (red dashed curve), for 
           those parameters which maximize the Holevo quantity, 
           for channel transmittivity $\eta=0.3$ (left) 
           and $\eta=0.8$ (right).} 
  \label{fig:Entanglement}
\end{figure}

In Fig.~\ref{fig:Entanglement} we plot both the 
average entanglement in the ensembles ${\cal G}_1$ (black full curve) 
and ${\cal G}_2$ (red dashed curve), for those 
parameters that solve the optimization problems (\ref{eq:max_chi_G1-2})  
and (\ref{eq:max_chi_G2}), respectively. 
As we can see, in the case of ${\cal G}_1$,  entanglement is more useful for 
poor channels (low values of $\eta$). For a given value of the 
transmissivity, the greater is the memory degree $\mu$ of the channel, 
the higher
is the amount of entanglement associated to the optimal ensemble ${\cal G}_1$. 
In the case of ${\cal G}_2$ we find that the presence of entanglement in the 
ensemble obeys a threshold behaviour. Actually the average entanglement 
(\ref{eq:entaglementG2}) 
vanishes if the population of the state $\ket{11}$ vanishes.
For ``good" quality channels ($\eta\gtrsim 0.7$), the entanglement 
associated to the optimal ensembles
behaves differently: ${\cal G}_1$ exhibits negligible average entanglement for
all values of the degree of memory, 
whereas ${\cal G}_2$ requires highly entangled states.

Finally, we want to comment on the $C_2$ capacity of a memoryless amplitude
damping channel ($\mu=0$). 
Since the Holevo quantity in general is not additive~\cite{hastings}
(and it has not been demonstrated to be additive for the amplitude damping 
channel), it is worth investigating whether entangled states may be 
useful to overcome the product state capacity $2\,C_{madc,1}$ 
(relative to two uses of a memoryless amplitude damping channel), namely
whether
\begin{equation}
%\max_{\{p_i,\ket{\psi_i}\}} \chi\big({\cal E}_0\otimes{\cal E}_0,
\max_{\{p_i,\ket{\psi_i}\}} \chi\big({\cal E}\otimes{\cal E},
\{p_i,\ket{\psi_i}\}\big)\underbrace{>}_? 2\,C_{{madc,1}},
\label{eq:C2}
\end{equation}
where $\{p_i,\ket{\psi_i}\}$ is a generic quantum ensemble in the Hilbert 
space of two qubits, and ${\cal E}$ is the single-qubit amplitude damping channel.
The answer to this question requires the optimization
in the left member of (\ref{eq:C2}) for any possible ensemble of  the
form (\ref{eq:generic-input-state2}, \ref{eq:diagonal-density-operator2}),
which is a very difficult task. 
We can, nevertheless investigate the behaviour of the ensembles 
${\cal G}_1$ and ${\cal G}_2$.
By numerical analysis it turns out that the maximization of the Holevo 
quantity over the ensemble ${\cal G}_2$ (\ref{eq:max_chi_G2}) 
always returns a value smaller than $2\,C_{madc,1}$, while the maximization on 
the class ${\cal G}_1$ (\ref{eq:max_chi_G1-2}) returns the value 
$2\, C_{madc,1}$.

%-----------------------------------QUANTUM CAPACITY---------------------------
\section{Quantum Capacity}
\label{sec:quantum-capacity}
In this section we consider the quantum capacity for the amplitude damping 
channel with memory and derive bounds for it.
We recall that the quantum capacity $Q$ is defined 
as~\cite{lloyd,barnum,devetak}
\begin{equation}
Q \,=\, \lim_{n\to\infty} \frac{Q_n}{n},
\quad \quad
Q_n\,=\,\max_{\rho^{(n)}} I_c(\mathcal{E}_{\mu}^{\otimes n},\rho^{(n)}),
\label{qinfo}
\end{equation}
where $\rho^{(n)}$ is an input state for $n$ channel uses and 
\begin{equation}
I_c(\mathcal{E}_{\mu}^{\otimes n},\rho^{(n)})\,=\,
S\big(\mathcal{E}_{\mu}^{\otimes n}\big( \rho^{(n)} \big) \big)\,-\,
S_e(\mathcal{E}_{\mu}^{\otimes n},\rho^{(n)})
\label{coinfo}
\end{equation}
is the \textit{coherent information}~\cite{schumachernielsen}.
In Eq. (\ref{qinfo}) $S_e(\mathcal{E}_{\mu}^{\otimes n},\rho^{(n)})$ is
the \textit{entropy exchange}~\cite{schumacher}, defined as 
\begin{equation}
S_e(\mathcal{E}_{\mu}^{\otimes n},\rho^{(n)})
=S\big[\big(\mathcal{I}\otimes \mathcal{E}_{\mu}^{\otimes n}\big)
\big(|\Psi\rangle\langle\Psi\big)\big],
\end{equation}
where $|\Psi\rangle$ is any purification of $\rho^{(n)}$, 
namely $\rho^{(n)}=\mathrm{Tr}_{\textsf R} |\Psi\rangle\langle\Psi|$ with
${\textsf R}$ denoting a reference system that evolves trivially, according 
to the identity superoperator $\mathcal{I}$.

In order to calculate the quantum capacity of the memory channel 
${\cal E}_\mu$ we need to deal with a unitary representation of this channel. 
This can be conveniently achieved by considering two external systems $E$ and 
$M$, the latter taking into account the degree of memory of the channel, as 
follows:
\begin{eqnarray}
&& \ket{00}^\textsf{S}\otimes\ket{00}^\textsf{E}\otimes\ket{0}^\textsf{M} \quad \longrightarrow \nonumber\\
  &&\hspace{1.5cm}\footnotesize \ket{00}^\textsf{S}\otimes\ket{00}^\textsf{E}\otimes \Big(\sqrt{1-\mu}\ket{0}^\textsf{M}+\sqrt{\mu}\ket{1}^\textsf{M}\Big), \normalsize
\nonumber\\
\nonumber\\
&& \ket{01}^\textsf{S}\otimes\ket{00}^\textsf{E}\otimes\ket{0}^\textsf{M} \quad \longrightarrow \quad 
    \sqrt{1-\mu}\,\Big(\sqrt{\eta}\ket{01}^\textsf{S}\otimes\ket{00}^\textsf{E}+\nonumber\\
     &&\hspace{1.5cm}       \sqrt{1-\eta}\ket{00}^\textsf{S}\otimes\ket{01}^\textsf{E}\Big)\otimes\ket{0}^\textsf{M} \,+\,
         \sqrt{\mu}\ket{01}^\textsf{S}\otimes\ket{00}^\textsf{E}\otimes\ket{1}^\textsf{M}, \nonumber\\
\nonumber\\
&& \ket{10}^\textsf{S}\otimes\ket{00}^\textsf{E}\otimes\ket{0}^\textsf{M} \quad \longrightarrow \quad
\sqrt{1-\mu}\,\Big(\sqrt{\eta}\ket{10}^\textsf{S}\otimes\ket{00}^\textsf{E}+\nonumber\\       
  &&\hspace{1.5cm} \sqrt{1-\eta}\ket{00}^\textsf{S}\otimes\ket{10}^\textsf{E}\Big)\otimes\ket{0}^\textsf{M} \,+\,\sqrt{\mu}\ket{10}^\textsf{S}\otimes\ket{00}^\textsf{E}\otimes\ket{1}^\textsf{M}, \nonumber\\
\nonumber\\
&& \ket{11}^\textsf{S}\otimes\ket{00}^\textsf{E}\otimes\ket{0}^\textsf{M} \quad \longrightarrow \nonumber\\
  &&\hspace{1.5cm} \sqrt{1-\mu}\Big[\eta\ket{11}^\textsf{S}\otimes\ket{00}^\textsf{E}\,+\,
\sqrt{\eta(1-\eta)}\Big(\ket{01}^\textsf{S}\otimes\ket{10}^\textsf{E}\,+\nonumber\\
&&\hspace{1.5cm}\ket{10}^\textsf{S}\otimes\ket{01}^\textsf{E}\Big)\,+\,(1-\eta)\ket{00}^\textsf{S}\otimes\ket{11}^\textsf{E}\Big]\otimes\ket{0}^\textsf{M}\,+\nonumber\\
&&\hspace{1.5cm}
\sqrt{\mu}\Big(\sqrt{\eta}\ket{11}^\textsf{S}\otimes\ket{00}^\textsf{E}\,+\,
		\sqrt{1-\eta}\ket{00}^\textsf{S}\otimes\ket{11}^\textsf{E}\Big)\otimes\ket{1}^\textsf{M}.\nonumber\\
\label{eq:unitary-representationEmu}
\end{eqnarray}
When the system $\textsf{S}$ is prepared in the generic pure state 
$\ket{\psi}$ the system $\textsf{SEM}$ state undergoes the transformation
\begin{eqnarray}
&&\hspace{-1cm} \ket{\psi^{\textsf{SEM}}}\,=\, \ket{\psi}^\textsf{S}\otimes\ket{000}^\textsf{EM}\,= \nonumber\\
&&\,a \ket{00}^\textsf{S}\otimes\ket{000}^\textsf{EM}\,+\,b\ket{01}^\textsf{S}\otimes\ket{000}^\textsf{EM}\,+\nonumber\\
&&\hspace{0cm} +                    
 \,c\ket{10}^\textsf{S}\otimes\ket{000}^\textsf{EM}\,+\,d\ket{11}^\textsf{S}\otimes\ket{000}^\textsf{EM} \qquad \longrightarrow\nonumber
\end{eqnarray}
\begin{eqnarray}
&&\hspace{-1.0cm} \ket{\psi^{\textsf{SEM}'}}\,=
a\sqrt{1-\mu}\ket{00}^\textsf{S}\otimes\ket{000}^\textsf{EM}+
                                     a\sqrt{\mu}\ket{00}^\textsf{S}\otimes\ket{001}^\textsf{EM}+
\nonumber\\
&& \hspace{0.5cm}b\sqrt{(1-\mu)\eta}\ket{01}^\textsf{S}\!\!\otimes\!\!\ket{000}^\textsf{EM}+
       b\sqrt{(1-\mu)(1-\eta)}\ket{00}^\textsf{S}\!\!\otimes\!\!\ket{010}^\textsf{EM}+\nonumber\\
&&\hspace{1.1cm}
        b\sqrt{\mu}\ket{01}^\textsf{S}\!\!\otimes\!\!\ket{001}^\textsf{EM}+\nonumber\\
&& \hspace{0.5cm}c\sqrt{(1-\mu)\eta}\ket{10}^\textsf{S}\!\!\otimes\!\!\ket{000}^\textsf{EM}+
       c\sqrt{(1-\mu)(1-\eta)}\ket{00}^\textsf{S}\!\!\otimes\!\!\ket{100}^\textsf{EM}+\nonumber\\
&&\hspace{1.1cm}
        c\sqrt{\mu}\ket{10}^\textsf{S}\!\!\otimes\!\!\ket{001}^\textsf{EM}+\nonumber\\
&& \hspace{0.5cm}d\sqrt{(1-\mu)}\eta\ket{11}^\textsf{S}\!\!\otimes\!\!\ket{000}^\textsf{EM}+\nonumber\\
&&\hspace{1.1cm}
       d\sqrt{(1-\mu)\eta(1-\eta)}\big(\ket{01}^\textsf{S}\!\!\otimes\!\!\ket{100}^\textsf{EM}+
\ket{10}^\textsf{S}\!\!\otimes\!\!\ket{010}^\textsf{EM}\big)+\nonumber\\
&&\hspace{1.1cm}
        d\sqrt{1-\mu}(1-\eta)\ket{00}^\textsf{S}\!\!\otimes\!\!\ket{110}^\textsf{EM}+\nonumber\\
&&\hspace{1.1cm}
        d\sqrt{\mu\eta}\ket{11}^\textsf{S}\!\!\otimes\!\!\ket{001}^\textsf{EM}+
 d\sqrt{\mu(1-\eta)}\ket{00}^\textsf{S}\!\!\otimes\!\!\ket{111}^\textsf{EM}.
\label{eq:pure-state-transformation-arbitrarymemory}
\end{eqnarray}
%\normalsize
>From equation (\ref{eq:pure-state-transformation-arbitrarymemory}) it is 
possible to obtain the expressions for the final state of the system, 
$\rho'={\cal E}_\mu(\rho)\equiv\rho^{\textsf{S}'}=\Tr_{\textsf{EM}}\big[\ketbra{\psi^{\textsf{SEM}'}}{\psi^{\textsf{SEM}'}}\big]$, 
and of the environment, 
$\rho^{\textsf{EM}'}=\Tr_{\textsf{S}}\big[\ketbra{\psi^{\textsf{SEM}'}}{\psi^{\textsf{SEM}'}}\big]$. 
We report their explicit form in the appendix
\ref{appx-sec:1}, see equations (\ref{eq-appx:rhoSout}) and 
(\ref{eq-appx:rhoEMout}).

%Even if 
The two extreme cases of memoryless ($\mu=0$) and full memory ($\mu=1$)
amplitude damping channels have been shown to be degradable 
\cite{giovannetti,MADC2013}, so that the regularization 
$n\to\infty$ in Eq. (\ref{qinfo}) is not necessary~\cite{degradable} 
and the quantum capacity is given by the single-shot formula, $Q=Q_1$. 
On the other hand, there is no evidence that degradability 
holds for the general case of partial memory. 
To hand the regularization formula in Eq. (\ref{qinfo}) is a hard task,
therefore we restrict to the computation of upper and lower bounds for the 
quantum capacity.

\subsection{An upper bound for $Q({\cal E}_\mu)$}

Since the channel ${\cal E}_\mu$ is a convex combination of 
the degradable channels ${\cal E}_0$ and ${\cal E}_m$,
according to
%${\cal E}_\mu(\rho)=(1-\mu){\cal E}_0(\rho)\,+\,\mu{\cal E}_m(\rho)$
Eq. (\ref{eq:model}), its quantum capacity is upper bounded 
by~\cite{smith-smolin07}
\begin{equation}
Q_{upb}\,=\,(1-\mu)Q({\cal E}_0)\,+\,\mu\, Q({\cal E}_m).
\label{eq:Smithsmolin-inequality} 
\end{equation}
This expression is easy to evaluate, since $Q({\cal E}_0)$ is known
from Ref~\cite{giovannetti}, and $Q({\cal E}_m)$ is known from
Ref~\cite{MADC2013}.

\subsection{A lower bound for $Q({\cal E}_\mu)$}

Here we use the ``single-letter'' formula $Q_1$, namely
\begin{equation}
  Q_1({\cal E}_\mu)\,=\,\max_{\rho}\,I_c({\cal E}_\mu,\rho),
\label{eq:QuantumCapacity1}
\end{equation}
where $\rho$ belongs to the Hilbert space corresponding to a single use 
of channel ${\cal E}_\mu$. The coherent information is then given by
\begin{eqnarray}
  I_c({\cal E}_\mu,\rho)=\,S({\cal E}_\mu(\rho))\,-S_e({\cal E}_\mu,\rho)=
S(\rho')\,-\,S(\rho^{\textsf{EM}'}),
  \label{eq:CoherentInformationArbitraryMemory}
\end{eqnarray}
where $S_e({\cal E}_\mu,\rho)\,=\,S(\rho^{\textsf{EM}'})$ 
is the entropy exchange related to ${\cal E}_\mu$~\cite{schumachernielsen}.

%Here $\rho\equiv\rho^\textsf{S}$ is a generic input state for the channel ${\cal E}_\mu$,
%$\rho'\equiv\rho^\textsf{S'}={\cal E}_\mu(\rho^{\textsf{S}})$ and $\rho^{\textsf{EM}'}$ are given by 
%(\ref{eq-appx:rhoSout}) and (\ref{eq-appx:rhoEMout}).
%
Since we do not know whether the coherent information of ${\cal E}_\mu$
is concave, we cannot simplify the form of the optimal input state by the
argument followed in the previous section for the Holevo quantity.
As far as we know, the concavity  holds for $I_c({\cal E}_\mu,\rho)$ 
only in the cases $\mu=0,1$.
For the generic case of $\mu\neq 0,1$ one should then 
try to maximize the coherent information 
(\ref{eq:CoherentInformationArbitraryMemory})
with respect to all possible input states $\rho^\textsf{S}$. 
This task is a hard task since it involves a maximization with respect to 
15 real parameters. 
We will then focus on a simpler task, by optimizing the coherent 
information (\ref{eq:CoherentInformationArbitraryMemory}) with respect 
to a diagonal input state
\begin{equation}
\hspace{-6.2cm}\rho = \left(
\begin{array}{cccc}
  \alpha & 0 & 0 & 0  \\
  0 & \beta & 0 & 0  \\
  0 & 0 & \gamma & 0  \\
  0 & 0 & 0 & \delta  \\
  \end{array} \right).
\label{eq:diagonal-density-operator-last} 
\end{equation}
This choice ensures that for $\mu=0$ and $\mu=1$, the corresponding 
bound gives the quantum capacity of the memoryless and of the full-memory 
channel, respectively, since the optimal input is a diagonal one for both 
channels, as shown in Ref.~\cite{giovannetti} and Ref.~\cite{MADC2013}.
The corresponding output density operators for the system $\textsf{S}$ and 
the environment $\textsf{ME}$ can be derived from equations 
(\ref{eq-appx:rhoSout}) and (\ref{eq-appx:rhoEMout}), and are shown below:
\begin{eqnarray}
 \hspace{-0.2cm}\rho' = \left(
  \begin{array}{cccc}
  \rho^{\textsf{S}'}_{00,00} & 0 & 0 & 0  \\
  0 & \rho^{\textsf{S}'}_{01,01} & 0 & 0  \\
  0 & 0 & \rho^{\textsf{S}'}_{10,10} & 0   \\
  0 & 0 & 0 & \rho^{\textsf{S}'}_{11,11}\\
  \end{array} \right),\label{eq:OutputDensityMatrixSystem-diag} \\ \nonumber
\end{eqnarray}
\begin{eqnarray}
 \hspace{-0.2cm}\rho^{\textsf{EM}'} = \left(
  \begin{array}{cccccccc}
  \rho^{\textsf{EM}'}_{000,000} & \rho^{\textsf{EM}'}_{000,001} & 0 & 0 & 0 & 0 & 0 & 0  \\
  \rho^{\textsf{EM}'}_{001,000} & \rho^{\textsf{EM}'}_{001,001} & 0 & 0 & 0 & 0 & 0 & 0  \\
  0 & 0 & \rho^{\textsf{EM}'}_{010,010} & 0  & 0 & 0 & 0 & 0 \\
  0 & 0 & 0 & 0 & 0 & 0 & 0 & 0 \\
  0 & 0 & 0 & 0 & \rho^{\textsf{EM}'}_{100,100} & 0 & 0 & 0 \\
  0 & 0 & 0 & 0 & 0 & 0 & 0 & 0 \\
  0 & 0 & 0 & 0 & 0 & 0 & \rho^{\textsf{EM}'}_{110,110} & \rho^{\textsf{EM}'}_{110,111} \\
  0 & 0 & 0 & 0 & 0 & 0 & \rho^{\textsf{EM}'}_{111,110} & \rho^{\textsf{EM}'}_{111,111} \\
  \end{array} \right), 
\label{eq:OutputDensityMatrixEnvironment-diag}
\\ \nonumber
\end{eqnarray}
\normalsize
where the matrix elements are reported in the appendix \ref{appx-sec:1} in
Eqs. (\ref{eq-appx:rhoSout}) and (\ref{eq-appx:rhoEMout}). 
Our lower bound for the quantum capacity of the channel 
${\cal E}_\mu$ is given by
\begin{equation} 
 Q_{lwb}=\max_{\alpha, \beta, \gamma, \delta} 
  \{I_c({\cal E}_\mu,\rho),0\}\,=\max_{\alpha, \beta, \gamma, \delta}\, 
  \Big\{\Big[S(\rho')\,-\,S(\rho^{\textsf{EM}'})\Big],0\Big\},
\label{eq:lwb-quantumcapacity-mu}
\end{equation}
whera $\alpha,\beta,\gamma,\delta\in [0,1]$, $\alpha+\beta+\gamma+\delta=1$, 
$\rho'$ and $\rho^{\textsf{EM}'}$ are given by
(\ref{eq:OutputDensityMatrixSystem-diag}) and 
(\ref{eq:OutputDensityMatrixEnvironment-diag}), respectively.
We solved the optimization problem (\ref{eq:lwb-quantumcapacity-mu}) numerically. 
The obtained results are reported in the following subsection.

\subsection{Numerical results}

\begin{figure}[h!]
  \begin{center}
  \includegraphics[width=13.cm]{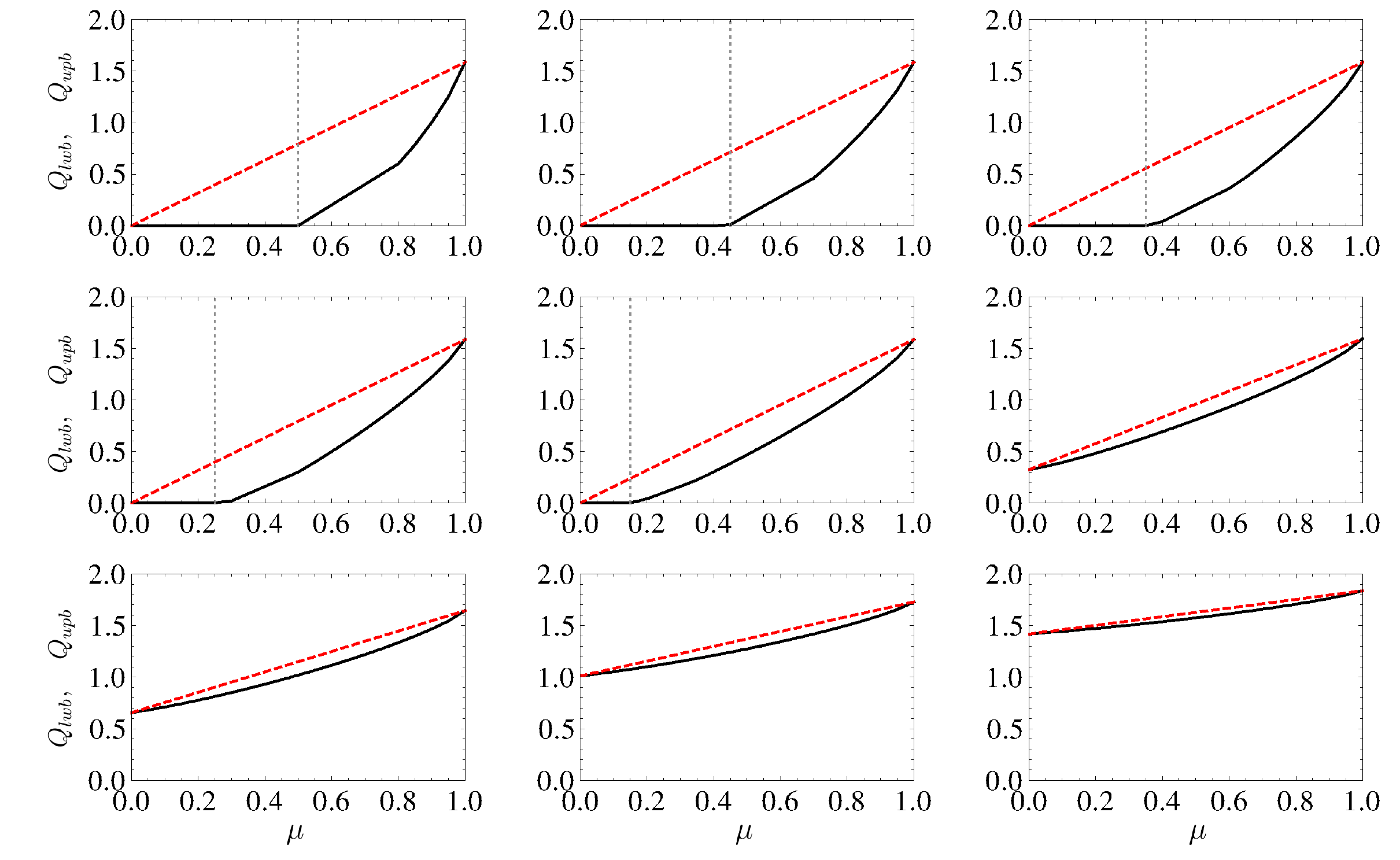}
  \end{center}
  \caption{(color online) Upper bound $Q_{upb}$ (\ref{eq:Smithsmolin-inequality}) (red dashed curve)
           and lower bound $Q_{lwb}$ (\ref{eq:lwb-quantumcapacity-mu}) (black full curve)
           for the quantum capacity of the channel ${\cal E}_\mu$.
           Different plots refer to  
           different channel transmissivities: from left to right, 
           $\eta=0,0.1,0.2$ (top row),
           $0.3,0.4,0.6$ (middle row), and $0.7,0.8,0.9$. 
           The dashed gray line signals the presence of a threshold $\bar{\mu}_{th}$:
           for values of the channel degree of memory $\mu\le\bar{\mu}_{th}$, 
           the lower bound (\ref{eq:lwb-quantumcapacity-mu})
           vanishes.}
  \label{fig:GeneralChannel-QuantumCapacity-mu}
\end{figure}

In Fig.~\ref{fig:GeneralChannel-QuantumCapacity-mu} we plot the bounds
(\ref{eq:Smithsmolin-inequality}) and (\ref{eq:lwb-quantumcapacity-mu}) as functions
of the memory degree $\mu$, for different values of the transmissivity parameter $\eta$.
We first notice that the lower bound (\ref{eq:lwb-quantumcapacity-mu}) 
exhibits a threshold value $\bar{\mu}_{th}$.
%Note that for $\eta=0.5$ the threshold value for having $Q_{lwb}>0$ is just $\bar{\mu}_{th}=0$. 
Indeed for $\mu\le\bar{\mu}_{th}$ we have that $Q_{lwb}\,=\,0$.
This threshold depends on the
channel transmissivity $\eta$, and it is only present for $\eta\le 0.5$. 
This is not  too surprising, since ${\cal E}_\mu$
is a convex combination of two channels and one of them, i.e. the memoryless
channel, has a vanishing quantum capacity for $\eta\le 0.5$. 
We would like to point out that for $\eta>0.5$ the chosen upper 
(\ref{eq:Smithsmolin-inequality})  
and lower bounds (\ref{eq:lwb-quantumcapacity-mu})
give good estimations of the quantum capacity for ${\cal E}_\mu$, since the 
corresponding values are close to each other, as one can see
from Fig.~\ref{fig:GeneralChannel-QuantumCapacity-mu}.

\begin{figure}[h!]
  \begin{center}
  \includegraphics[width=10cm]{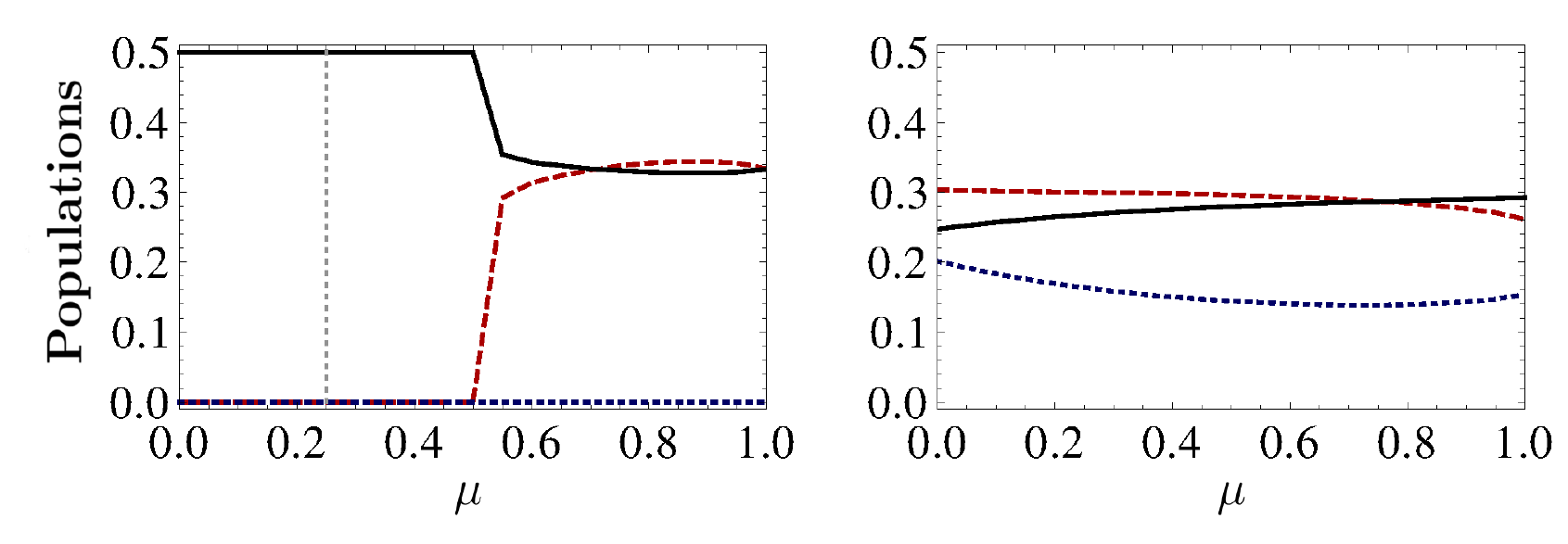}
  \end{center}
  \caption{(color online) Populations $\alpha$ (long-dashed red curve), 
           $\beta=\gamma$ (black curve),  and
           $\delta$ (dashed blue curve) which solve the maximization problem 
           (\ref{eq:lwb-quantumcapacity-mu}),
           for $\eta=0.3$ (left) and $\eta=0.8$ (right). 
           %One can notice that the maximization (\ref{eq:lwb-quantumcapacity-mu}) 
           %returns equal values for the populations of the states 
           %$\ket{01}$ and $\ket{10}$ ($\beta=\gamma$).
           The dashed gray curve signals the presence of a threshold $\bar{\mu}_{th}$:
           for values of the channel degree of memory $\mu\le\bar{\mu}_{th}$, 
           the maximum of the coherent information 
           (\ref{eq:CoherentInformationArbitraryMemory})
           with respect to the input (\ref{eq:diagonal-density-operator-last}) 
           is smaller than or equal to 0.}
  \label{fig:GeneralChannel-QuantumCapacity-mu-coefficients}
\end{figure}

In Fig.~\ref{fig:GeneralChannel-QuantumCapacity-mu-coefficients} we plot
the values of the populations $\alpha,\,\beta,\,\gamma,\,\delta$ 
(\ref{eq:diagonal-density-operator-last}),
which solve the maximization problem~(\ref{eq:lwb-quantumcapacity-mu}).
We notice that the maximization problem~(\ref{eq:lwb-quantumcapacity-mu}) 
returns equal populations for the states $\ket{01}$ and $\ket{10}$, 
$\beta=\gamma$.
For low values of transmissivity ($\eta\le0.5$) the state 
$\ket{11}$ is not populated. This can be explained by some considerations.
First, we notice that the state $\ket{11}$
is the one which experiences the strongest noise 
(greatest damping rates), see
the Kraus operators $A_0$ in Eqs.~(\ref{eq:memoryless}) and 
$B_0$ in Eqs.~(\ref{eq:memory-Kraus-Operators}). 
Moreover, we remind that the channel ${\cal E}_\mu$ is a convex combination
of the memoryless channel ${\cal E}_0$ and the full memory channel 
${\cal E}_1$.
For $\eta\le0.5$, %the channel ${\cal E}_0$ has vanishing quantum capacity~\cite{giovannetti},
only the channel ${\cal E}_1$ has a non vanishing quantum capacity
\cite{MADC2013} and the optimal ensemble which maximizes the coherent 
information of 
${\cal E}_1$ is a diagonal one (\ref{eq:diagonal-density-operator-last}), 
with vanishing populations $\delta$ (for $\eta\le 0.5$), 
as reported in Ref.~\cite{MADC2013}.

%---------------ENTANGLEMENT ASSISTED CLASSICAL CAPACITY----------------------

\section{Classical Entanglement-Assisted Capacity}
\label{sec:classical-Entanglement-Assisted Capacity}

In this section we compute the entanglement-assisted classical capacity $C_E$,
which gives the maximum amount of classical information
that can be reliably transmitted down the channel per channel use, provided 
the sender and the receiver share an infinite amount of prior entanglement.
It is given by~\cite{bennett1999,bennett-shor}
\begin{equation}
C_E \,=\max_{\rho} I(\mathcal{E}_{\mu},\rho),
\label{CECapacity}
\end{equation}
where the maximization is performed over 
the input state $\rho$ for a single use of the channel
$\mathcal{E}_{\mu}$ and 
\begin{equation}
I(\mathcal{E}_{\mu},\rho) \,= S(\rho) + 
I_c (\mathcal{E}_{\mu},\rho).
\end{equation}
The subadditivity of $I$~\cite{adami-Cerf} guarantees 
that no regularization as in (\ref{qinfo}) is required
to obtain $C_E$.  

%\subsection{Maximization of $I$}
%\label{sec:Maximization-I}

By exploiting the concavity of $I$~\cite{adami-Cerf} and the covariance 
properties of the channel, following similar arguments as the ones reported 
in sect. 
\ref{sec:max-chi}, we can prove that the state $\rho$ maximizing $I$ is 
diagonal with the same populations for the states $\ket{01}$ and 
$\ket{10}$, as in Eq. (\ref{eq:diagonal-density-operator2}).
Therefore
\begin{equation} 
 C_E=\max_{\alpha, \beta,\delta} 
  I({\cal E}_\mu,\rho),\,=\max_{\alpha, \beta,  \delta}\, 
  [S(\rho^{\textsf{S}})\,+\,S(\rho^{\textsf{S}'})\,-\,S(\rho^{\textsf{EM}'})\Big].
\label{eq:cea-mu}
\end{equation}

\begin{figure}[h!]
  \begin{center}
  \includegraphics[width=9cm]{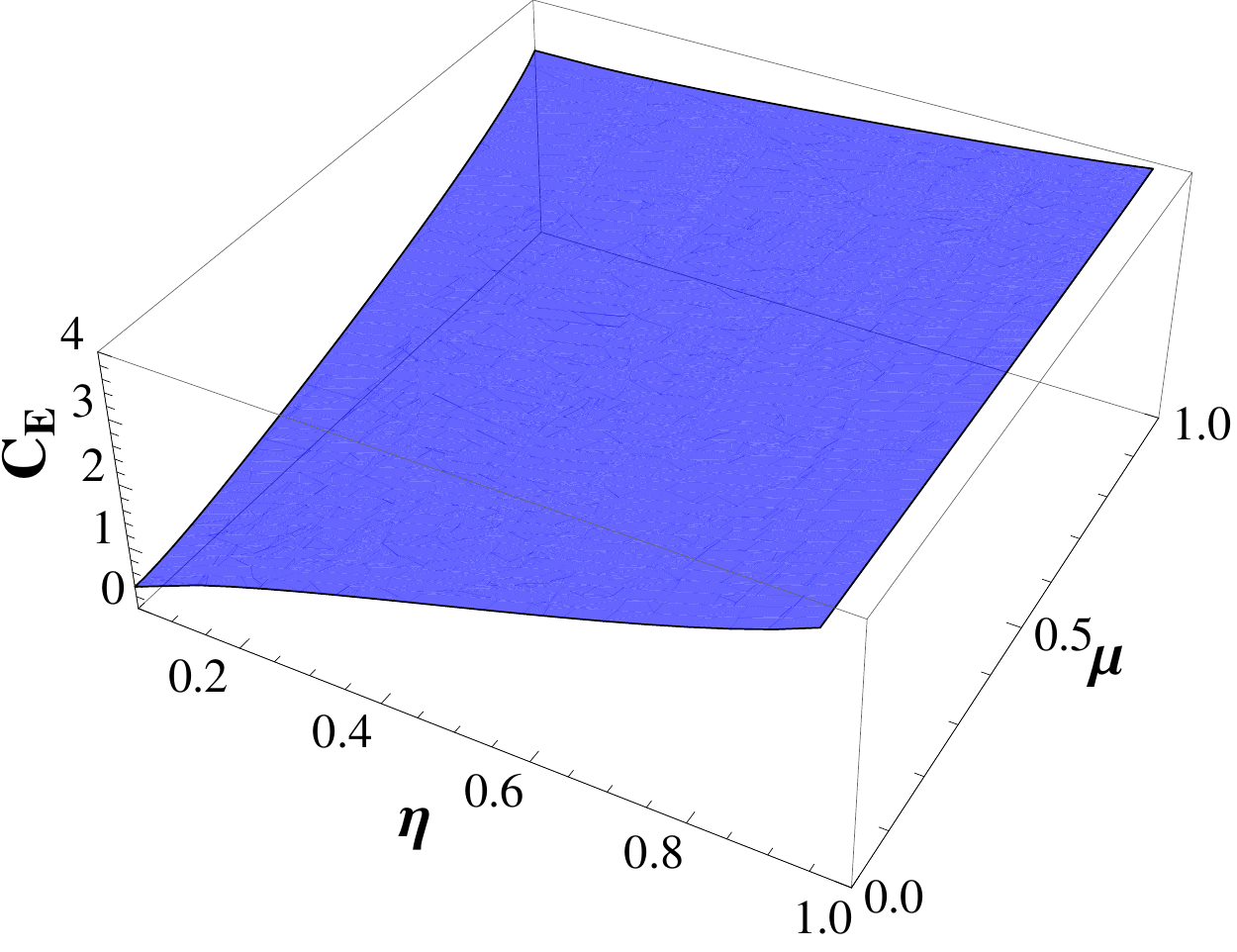}
  \end{center}
  \caption{(color online) Entanglement-assisted classical capacity of the channel ${\cal E}_\mu$,
           as a function of the transmittivity $\eta$ and of the degree of memory $\mu$.}
  \label{fig:GeneralChannel-ClassicalEntAssCapacity-mu}
\end{figure}

The numerical results achieved by maximization of the above expression are
reported in Fig.~\ref{fig:GeneralChannel-ClassicalEntAssCapacity-mu}.
As we can see, for any fixed value of $\eta$ the entanglement assisted capacity
is an increasing function of the degree of memory. Therefore, memory effects
are beneficial to improve the performance of the channel.
In particular, for $\eta=0$ we have a qualitative similar behaviour as the 
classical capacity. Actually, we can see that $C_E$ is vanishing 
in the memoryless case, but it is always nonzero as soon as the channel
has some memory, achieving the maximum value 3 for the full memory case. 

\begin{figure}[h!]
  \begin{center}
  \includegraphics[width=10cm]{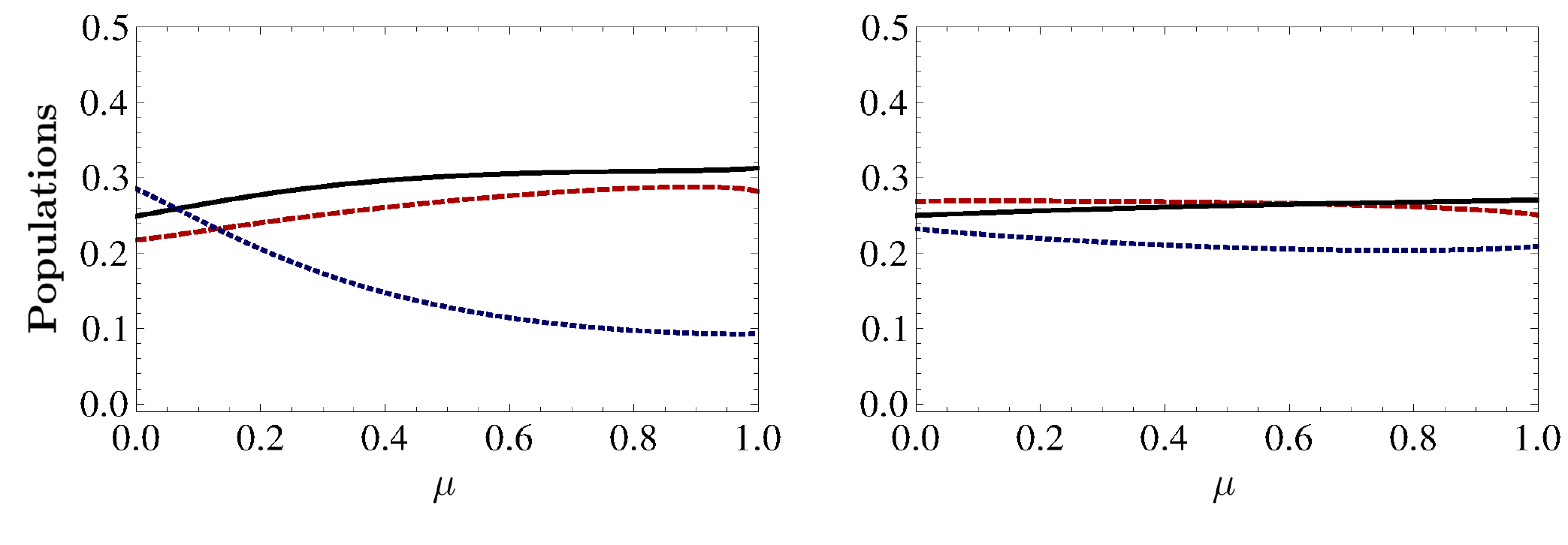}
  \end{center}
  \caption{(color online) Coefficients  $\alpha$ (red long-dashed curve), 
           $\beta=\gamma$ (black full curve),
           $\delta$ (blue dashed curve) which solve the maximization problem 
           (\ref{eq:cea-mu}), for $\eta=0.3$ (left) and $\eta=0.8$ (right).}
  \label{fig:GeneralChannel-ClassicalEntAssCapacity-PoPmu}
\end{figure}

In Fig.~\ref{fig:GeneralChannel-ClassicalEntAssCapacity-PoPmu}
we plot the populations of the state (\ref{eq:diagonal-density-operator2})
which solve the maximization problem (\ref{eq:cea-mu}).

\section{Conclusions}
\label{sec:conclusions}

In this work we have studied the performance of an amplitude 
damping channel with memory acting on a two qubits system. We considered 
a general noise model with arbitrary degree of memory, that includes
the memoryless amplitude damping channel and the full memory amplitude
damping channel as particular cases.
We have analysed three types of scenarios for information transmission.
We have first considered the transmission of classical information and 
have derived lower bounds on the classical channel capacity for a single use 
of the channel by numerical optimisation of the Holevo quantity for two 
significant types of input ensembles. We have then considered the case of 
quantum information and computed upper and lower bounds for the quantum 
capacity. We emphasized that for high values of the channel transmissivity 
it turns out that
the upper and lower bounds are quite close to each other, thus providing
a good estimate of the quantum channel capacity.
Finally, we computed the entanglement assisted classical channel capacity
numerically for any value of the channel transmissivity $\eta$ and degree of 
memory $\mu$.

\begin{acknowledgments}
A. D'A. and G. Falci acknowledge support from Centro Siciliano di Fisica 
Nucleare e Struttura della Materia (CSFNSM) Catania.
G.B. acknowledges the support by MIUR-PRIN project
``Collective quantum phenomena: From strongly correlated systems to
quantum simulators''.
\end{acknowledgments}

\appendix

\section{Coherent Information for an amplitude damping channel with arbitrary 
degree of memory}
\subsection{Expressions for $\rho^{\textsf{S}'}$ and $\rho^{\textsf{EM}'}$}
\label{appx-sec:1}

We describe a generic initial state of the system by the density operator
\begin{equation}
\rho^{\textsf{S}}  =\left(
\begin{array}{cccc}
  \alpha     & \kappa            & \lambda               & \xi    \\
  \kappa^*   & \beta             & \nu                   & o      \\
  \lambda^*  & \nu^*             & \gamma                & \pi    \\
  \xi^*      & o^*               & \pi^*                 & \delta \\
  \end{array} \right).
\label{eq:output-genericstate-2} 
\end{equation}
The output state of the system $\textsf{S}$ and of the environment 
$\textsf{EM}$ 
can be derived from equation 
(\ref{eq:pure-state-transformation-arbitrarymemory}). 
We report only the upper triangular part of $\rho^{\textsf{S}'}$ and
$\rho^{\textsf{EM}'}$, since any density operator matrix is an Hermitian 
matrix.

\subsubsection{Matrix $\rho^{\textsf{S}'}$}

In the basis $\{\ket{ij}^\textsf{S}\},\,\, i,j \in \{0,1\}$, 
the $\rho^{\textsf{S}'}$ matrix elements are given by (we set
$\rho^{\textsf{S}'}_{ij,i'j'}\equiv \, ^\textsf{S}\!\bra{ij}
\rho^{\textsf{S}'}\ket{i'j'}^\textsf{S}$)
\begin{eqnarray}
&& \rho^{\textsf{S}'}_{00,00}=(1-\mu)[\alpha+(1-\eta)(\beta+\gamma)+ 
                     (1-\eta)^2\delta]+\mu[\alpha +(1-\eta)\delta], \nonumber\\
&& \rho^{\textsf{S}'}_{00,01}=(1-\mu)[\sqrt{\eta}\kappa+\sqrt{\eta}(1-\eta)\pi]+\mu \kappa,\nonumber\\
&& \rho^{\textsf{S}'}_{00,10}=(1-\mu)[\sqrt{\eta}\lambda+\sqrt{\eta}(1-\eta)o]+\mu \lambda,\nonumber\\
&& \rho^{\textsf{S}'}_{00,11}=[(1-\mu)\,\eta+\mu\, \sqrt{\eta}]\,\xi,\nonumber\\
&& \rho^{\textsf{S}'}_{01,01}=(1-\mu)[\eta\beta+ 
            \eta(1-\eta)\delta]+\mu\beta, \nonumber\\
&& \rho^{\textsf{S}'}_{01,10}=[(1-\mu)\,\eta+\mu]\,\nu,\nonumber\\
&& \rho^{\textsf{S}'}_{01,11}=[(1-\mu)\,\eta^\frac{3}{2}+\mu\sqrt{\eta}]\,o,\nonumber\\
&& \rho^{\textsf{S}'}_{10,10}=(1-\mu)[\eta\gamma+ 
   \eta(1-\eta)\delta]+\mu\gamma,  \nonumber\\
&& \rho^{\textsf{S}'}_{10,11}=[(1-\mu)\,\eta^\frac{3}{2}+\mu\sqrt{\eta}]\,\pi,\nonumber\\
&& \rho^{S'}_{11,11}=(1-\mu)\eta^2\delta+ \mu\eta\delta. 
\label{eq-appx:rhoSout}
\end{eqnarray}

\subsubsection{Matrix $\rho^{\textsf{EM}'}$}

The elements of the output environment density matrix $\rho^{\textsf{EM}'}$
in the basis $\{\ket{ijk}^\textsf{EM}\},\,\,i,j,k \in \{0,1\}$, are given 
by (we set $\rho^{S'}_{ijk,i'j'k'}\equiv \, ^\textsf{EM}\!\bra{ijk}\rho^{S'}
\ket{i'j'k'}^\textsf{EM}$)
\begin{eqnarray}
&& \rho^{\textsf{EM}'}_{000,000}=(1-\mu)\,[\alpha+\eta(\beta+\gamma)+ 
                     \eta^2\delta], \nonumber\\
&& \rho^{\textsf{EM}'}_{000,001}=\sqrt{\mu(1-\mu)}\,[\alpha+\sqrt{\eta}(\beta+\gamma)+\eta^\frac{3}{2}\delta],\nonumber\\
&& \rho^{\textsf{EM}'}_{000,010}=(1-\mu)\sqrt{1-\eta}\,(\kappa+\eta\,\pi),\nonumber\\
&& \rho^{\textsf{EM}'}_{000,011}=0,\nonumber\\
&& \rho^{\textsf{EM}'}_{000,100}=(1-\mu)\sqrt{1-\eta}\,(\lambda+\eta\, o),\nonumber\\
&& \rho^{\textsf{EM}'}_{000,101}=0,\nonumber\\
&& \rho^{\textsf{EM}'}_{000,110}=(1-\mu)(1-\eta)\,\xi,\nonumber\\
&& \rho^{\textsf{EM}'}_{000,111}=\sqrt{\mu(1-\mu)(1-\eta)}\,\xi,\nonumber
\end{eqnarray}
\begin{eqnarray}
&&\rho^{\textsf{EM}'}_{001,001}=\mu[1-(1-\eta)\delta], \nonumber\\
&&\rho^{\textsf{EM}'}_{001,010}=\sqrt{\mu(1-\mu)(1-\eta)}\,(\kappa+\eta\,\pi),\nonumber\\
&&\rho^{\textsf{EM}'}_{001,011}=0,\nonumber\\
&&\rho^{\textsf{EM}'}_{001,100}=\sqrt{\mu(1-\mu)(1-\eta)}\,(\lambda+\eta\,o),\nonumber\\
&&\rho^{\textsf{EM}'}_{001,101}=0,\nonumber\\
&&\rho^{\textsf{EM}'}_{001,110}=\sqrt{\mu(1-\mu)}(1-\eta)\,\pi,\nonumber\\
&&\rho^{\textsf{EM}'}_{001,111}=\mu\sqrt{1-\eta}\,\pi,\nonumber
\end{eqnarray}
\begin{eqnarray}
&&\rho^{\textsf{EM}'}_{010,010}=(1-\mu)(1-\eta)\,(\beta+\eta\,\delta),\nonumber\\
&&\rho^{\textsf{EM}'}_{010,011}=0,\nonumber\\
&&\rho^{\textsf{EM}'}_{010,100}=(1-\mu)(1-\eta)\,\nu,\nonumber\\
&&\rho^{\textsf{EM}'}_{010,101}=0,\nonumber\\
&&\rho^{\textsf{EM}'}_{010,110}=(1-\mu)(1-\eta)^\frac{3}{2}\,o,\nonumber\\
&&\rho^{\textsf{EM}'}_{010,111}=\sqrt{\mu(1-\mu)}(1-\eta)\,o,\nonumber
\end{eqnarray}
\begin{eqnarray}
&&\rho^{\textsf{EM}'}_{011,ijk}=0\, \quad \forall i,j,k\,\in\{0,1\},\nonumber
\end{eqnarray}
\begin{eqnarray}
&&\rho^{\textsf{EM}'}_{100,100}=(1-\mu)(1-\eta)\,(\gamma+\eta\,\delta),\nonumber\\
&&\rho^{\textsf{EM}'}_{100,101}=0,\nonumber\\
&&\rho^{\textsf{EM}'}_{100,110}=(1-\mu)(1-\eta)^\frac{3}{2}\,\pi,\nonumber\\
&&\rho^{\textsf{EM}'}_{100,111}=\sqrt{\mu(1-\mu)}(1-\eta)\,\pi,\nonumber
\end{eqnarray}
\begin{eqnarray}
&&\rho^{\textsf{EM}'}_{101,ijk}=0\, \quad \forall i,j,k\,\in\{0,1\},\nonumber
\end{eqnarray}
\begin{eqnarray}
&&\rho^{\textsf{EM}'}_{110,110}=(1-\mu)(1-\eta)^2\delta,\nonumber\\
&&\rho^{\textsf{EM}'}_{110,111}=\sqrt{\mu(1-\mu)}(1-\eta)^\frac{3}{2}\,\delta,\nonumber
\end{eqnarray}
\begin{eqnarray}
&&\rho^{\textsf{EM}'}_{111,111}=\mu(1-\eta)\delta.
\label{eq-appx:rhoEMout}
\end{eqnarray}

\end{document}